\documentclass[prb,reprint,amssymb,showpacs,aps]{revtex4-1}
\usepackage{graphicx}
\usepackage{amsmath}
\usepackage{braket}
\usepackage{bm}
\usepackage{natbib}

\begin{document}

\title{Long-range Magnetic Order in Models for Rare Earth Quasicrystals}

\author{Stefanie Thiem}
\author{J. T. Chalker}
\affiliation{Theoretical Physics, University of Oxford, 1 Keble Road, Oxford OX1 3NP, United Kingdom}
\date{\today}

\begin{abstract}
We take a two-step theoretical approach to study magnetism of rare earth quasicrystals by considering Ising spins on quasiperiodic tilings, coupled via RKKY interactions. First, we compute RKKY interactions from a tight-binding Hamiltonian defined on the two-dimensional quasiperiodic tilings. We find that the magnetic interactions are frustrated and strongly dependent on the local environment. This results in the formation of clusters with strong bonds at certain patterns of the tilings that repeat quasiperiodically. Second, we examine the statistical mechanics of Ising spins with these RKKY interactions, using extensive Monte Carlo simulations. Although models that have frustrated interactions and lack translational invariance might be expected to display spin glass behaviour, we show that the spin system has a phase transition to low-temperature states with long-range quasiperiodic magnetic order. Additionally, we find that in some of the systems spin clusters can fluctuate much below the ordering temperature.
\end{abstract}

\pacs{75.50.Kj, 71.15.-m} 

\maketitle


Quasicrystals with their unusual atomic structure characterized by long-range order without a three-dimensional translational periodicity \cite{PhysRevLett.1984.Shechtman} are known to have rather exotic physical properties, and so far we lack a good theoretical understanding for many of them. For instance, the electronic properties of this material class include a pseudogap at the Fermi energy \cite{PhysRevLett.1991.Fujiwara, PhysRevLett.1992.Hafner} and multifractal (critical) wave functions \cite{PhysRevB.1987.Kohmoto} resulting in anomalous electronic transport \cite{PhysicalProperties.1999.Stadnik, JPhysJap.1987.Hiramoto}. An important question in this context is how these electronic properties influence the magnetism in quasicrystals.\cite{PhilMag.2008.Hippert, MagProp.2013.Stadnik}  

In general, one can distinguish two types of magnetic quasicrystals: those with magnetic moments at (i) transition metal or (ii) rare earth sites. In the first, moment formation is an important aspect of the theoretical problem since they appear only at a small fraction of sites.\cite{PhilMag.2008.Hippert} By contrast, the second class suggests a simpler description, with well-defined local moments at concentrations of 5-10\% interacting via long-range RKKY interactions which are mediated by the conduction electrons.\cite{PhysRevB.1999.Fisher,NatureMat.2013.Goldman} Examples of the latter class include the icosahedral i-ZnMgR and i-AgInR  materials\cite{PhysRevB.1999.Fisher, PhilMagLett.2002.Guo}, as well as decagonal d-ZnMgR materials \cite{PhilMagLett.1998.Sato} and the recently discovered binary phases\cite{NatureMat.2013.Goldman} i-RCd.

Here we use tight-binding models and Ising models to address this theoretical problem. We examine the form of the RKKY interactions based on a tight-binding Hamiltonian defined on two-dimensional quasiperiodic tilings (see Section \ref{sec:rkky}). We find that the coupling between pairs of sites depends not only on their distance but also varies strongly with the local environment on the tiling. Although we find ferromagnetic and antiferromagnetic bonds as in periodic systems, the magnetic interactions do not show a well-defined spatial period with a Fermi wave vector because quasicrystals do not have a Brillouin zone. Second, we study the consequences of these interactions (see Section \ref{sec:pt}), by taking them as exchange constants between Ising spins located at a fraction of sites on the tiling (those with a particular coordination number). As all systems show a combination of frustration and aperiodicity one might expect to observe a spin glass at low temperatures \cite{RevModPhys.1986.Binder}, and in other settings quasiperiodic systems are known to behave like random ones \cite{PhysRevLett.1982.Fishman}. Our results exclude canonical spin glass behaviour via the temperature dependence of the order parameter susceptibility, the overlap susceptibility, and the magnetic structure factor of the ground state. Instead we find that the ground state consists of repeating small clusters of spins, with strong interactions within each cluster and weaker couplings between different clusters. Analysing the magnetic structure factor we find that these systems show a transition to a state with long-range magnetic order at low temperatures.

While there has been extensive previous work studying spin models for magnetic quasicrystals, to the best of our knowledge none has used RKKY interactions computed from a quasiperiodic electron Hamiltonian. Moreover much of this previous work omits the frustration effects that are a natural consequence of long-range oscillatory RKKY interactions. Nearest neighbour exchange on a bipartite tiling necessarily leads in an Ising antiferromagnet \cite{JStatPhys.1986.Godreche} to a classical ground state with two-sublattice order. In Heisenberg models it is known that this order may survive quantum fluctuations \cite{PhysRevLett.2003.Wessel} or the inclusion of dipolar and further neighbour interactions. \cite{PhysRevLett.2003.Vedmedenko,PhysRevLett.2004.Vedmedenko,PhilMag.2006.Vedmedenko, JStatPhys.1986.Godreche} RKKY interactions with a form taken from periodic systems have been used in studies of Ising models for the Penrose tiling \cite{JNonCrys.2004.Matsuo} and for i-ZnMgHo models\cite{JMagMagMat.2002.Matsuo, PhilMag.2006.Matsuo}, yielding antiferromagnetic or ferrimagnetic ordered ground states. Neither of these models incorporates the expected unique coupling of the exotic electronic properties to the magnetism in quasicrystals. Our model is designed to address this aspect of the physics in a simple way.

In previous work (Ref.~\onlinecite{EPL.2015.Thiem}) we have studied the structure of the RKKY interactions for a limited number of systems. Here we present a much more thorough analysis and classification of quasiperiodic spin systems. An important improvement is a new algorithm for the computation of the RKKY interactions which has higher accuracy and is also applicable to systems with gaps in the density of states, in contrast to the previously employed continued fraction expansion. We use the new algorithm to compute the RKKY interactions for a large set of systems. This allows us to identify different classes of behaviour with respect to the structure of the magnetic ground state. The new results also provide a more detailed insight into the nature of low-temperature fluctuations. Finally, we apply finite size scaling to show that the critical behaviour is consistent with the two-dimensional Ising universality class.

\section{RKKY Interactions in Tight-Binding Models}
\label{sec:rkky}

\begin{figure}[t!]
    \centering
    \includegraphics[width=\columnwidth]{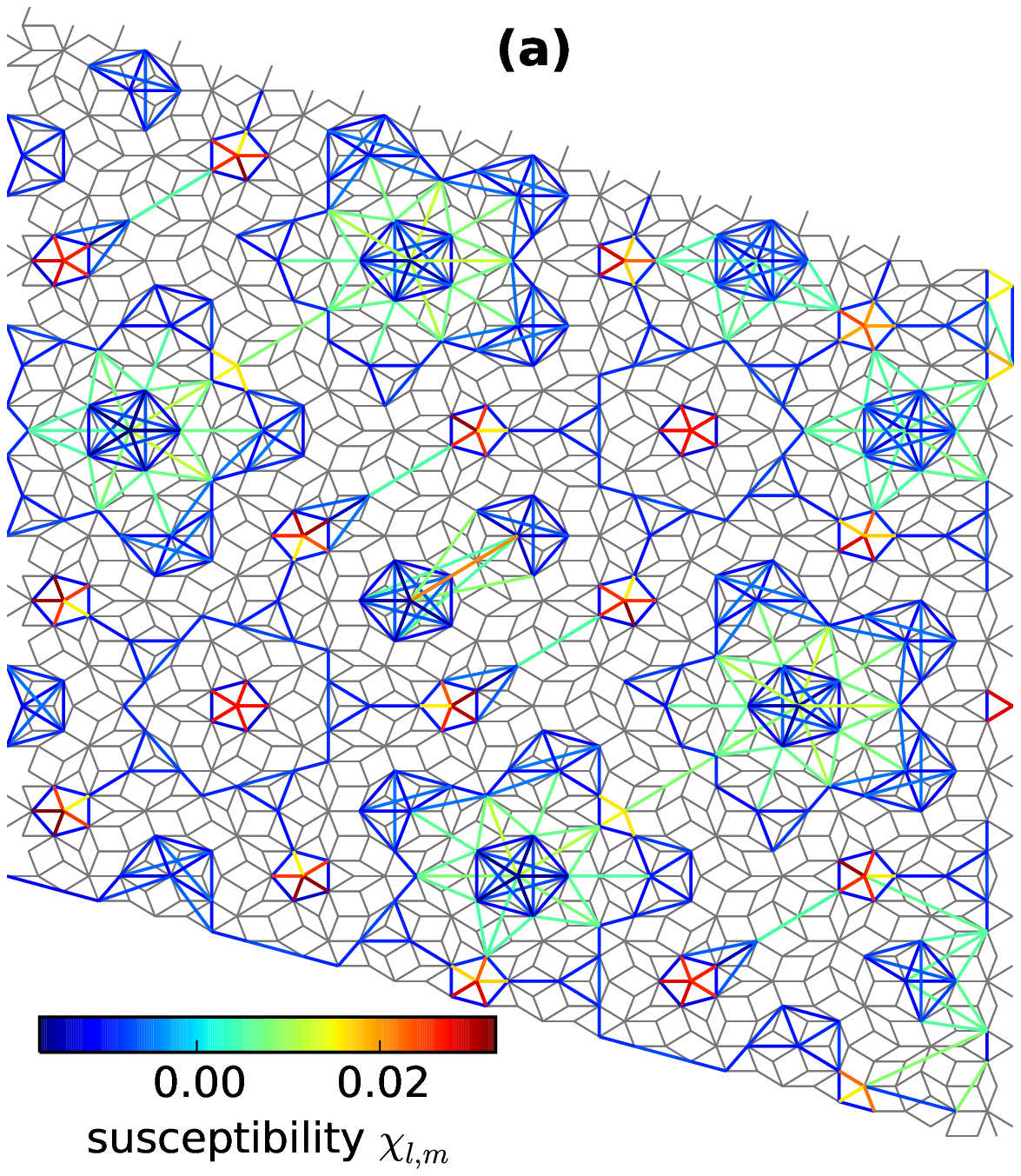}
    \includegraphics[width=\columnwidth]{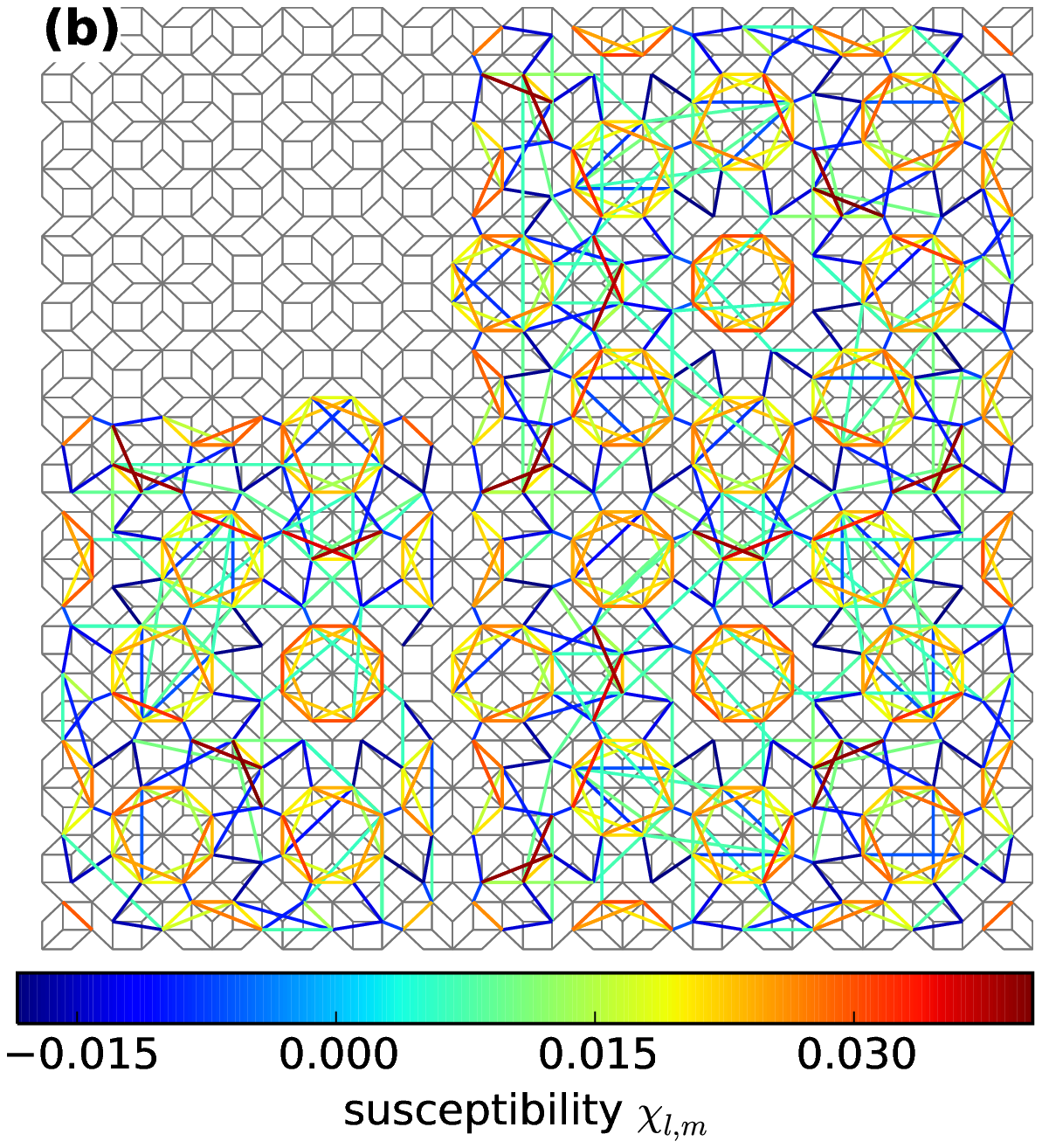}
  \caption{Tilings and RKKY interactions $\chi_{l,m}$ for magnetic moments at sites with coordination number $z$, coupling strength $\lambda=1$, and $\eta = 0.03$: (a) 5th approximant of the Penrose tiling, showing only interactions above the threshold  $|\chi_{l,m}| > 0.005$, $z=5$, and $E_\mathrm{F} = 0.07$; (b) 4th approximant of the Ammann Beenker tiling with $|\chi_{l,m}| > 0.006$, $z=4$, and $E_\mathrm{F} = 1.12$.}
\label{fig:tilings}
\end{figure}

Introducing Ising spins $\sigma_l$ at some of the sites $l$ of the tiling, the spin Hamiltonian has the form  $H_\mathrm{RKKY} = \lambda^2 \chi_{l,m} \sigma_l \sigma_m$. Here, $\lambda$ represents the coupling of the local moment to the conduction electrons. We model the conduction electrons by the tight-binding Hamiltonian $H_\textrm{el} = \sum_{\langle l,m \rangle} \ket{l}\bra{m}$ with one orbital per site and equal hopping amplitudes between all nearest neighbours of a quasiperiodic tiling (see Fig.~\ref{fig:tilings}).  The local susceptibility \cite{JPhysFrance.1994.Jagannathan}  for $T = 0$ and Fermi energy $E_\mathrm{F}$ is (for a review see Ref.~\onlinecite{Crystals.2013.Power})
\begin{equation}
    \label{equ:susceptibility}
    \chi_{l,m} = \frac{1}{\pi} \int_{-\infty}^{\infty} \Im\left[ G_{l,m} G_{m,l} \right] \mathrm{sign} (E - E_\mathrm{F}) \mathrm{d}E . 
\end{equation}
Here $G_{l,m} \equiv \bra{l} G \ket{m}$ is a matrix element of the retarded Green function $G\equiv [E+\imath0-H_\textrm{el}]^{-1}$ for the conduction electrons. Writing the Green functions in terms of the eigenstates as 
\begin{equation}
	G_{l,m} = \sum_{\alpha} \frac{\Phi_\alpha(l) \Phi_\alpha(m)}{E-E_\alpha+\imath0} \;,
\end{equation}
Eq.~\eqref{equ:susceptibility} evaluates to 
\begin{equation}
    \label{equ:susceptibility2}
    \chi_{l,m} = 2 \sum_{\alpha, \beta} \frac{\Phi_\alpha(l) \Phi_\alpha(m) \Phi_\beta(l) \Phi_\beta(m)}{E_\alpha - E_\beta} \;\mathrm{sign} (E_\mathrm{F} - E_\alpha) \;,
\end{equation}
where the sum runs over all eigenstates with either $E_\alpha < E_\mathrm{F} \le E_\beta$ or $E_\beta < E_\mathrm{F} \le E_\alpha$. This equation clearly  shows that the properties of the electronic eigenstates of the Hamiltonian $H_\textrm{el}$ have an important influence on the RKKY interactions. In the case of quasicrystals, this approach also naturally includes the multifractal structure of the electronic wave functions.\cite{JPhys.1998.Rieth,PhilosMag.2011.Trambly}

The numerical evaluation of Eq.~\eqref{equ:susceptibility2} leads to a divergence of the susceptibility if the Fermi energy $E_F$ is close to an energy eigenvalue $E_\alpha$. To overcome this problem we approximate the $\mathrm{sign}$-function in Eq.~\ref{equ:susceptibility} with $\tanh((E - E_\mathrm{F})/\eta)$, where $\eta$ can be interpreted as a temperature parameter. This yields a good approximation as long as the band width of the system $\gg \eta$. We use $\eta = 0.03$ for all calculations which is much smaller than the band width for the considered quasiperiodic systems. Re-evaluating Eq.~\ref{equ:susceptibility}, we can then rewrite Eq.~\ref{equ:susceptibility2} replacing the $\mathrm{sign} (E_\mathrm{F} - E_\alpha)$ with the expression 
\begin{equation}
	 \frac{ \sinh{\left(\frac{E_\alpha - E_\beta}{\eta}\right)} } {2 \cosh{\left(\frac{E_\alpha - E_\mathrm{F}}{\eta}\right)} \cosh{\left(\frac{E_\mathrm{F} - E_\beta}{\eta}\right)} } \;.
\end{equation}

Our approach is based on the direct diagonalization of the electron Hamiltonian $H_\textrm{el}$. Using fast numerical libraries (lapack) and parallelization we are able to compute the magnetic interactions for systems with up to $10^4$ sites. Our method is more accurate than the continued fraction expansion, which has been employed previously for quasiperiodic tilings. \cite{JPhysFrance.1994.Jagannathan, PhysRevB.1987.Kumar, EPL.2015.Thiem}

The method can be applied to general tight-binding systems. For a particular model we need to specify three features: (i) the tiling; (ii) the positions of the spins on the tiling; and (iii) the value of the Fermi energy $E_\mathrm{F}$. In this paper we study two different quasiperiodic systems. \\
(A) Penrose tiling (see Fig.~\ref{fig:tilings}a): This tilings can be used to describe the structure of decagonal quasicrystals. \cite{ActaCrys.1988.Yamamoto} It has 5 different local environments with 3 to 7 nearest neighbours. We consider the 4th to 6th approximants with 521, 1364, and 3571 sites respectively.\\
(B) Ammann Beenker tiling (see Fig.~\ref{fig:tilings}b): This tiling models the structure of octagonal quasicrystals \cite{DCGeom.1992.Ammann} and possesses 6 different local environments with 3 to 8 nearest neighbours. We consider the 4th and 5th approximant with 1393 and 8119 sites respectively.\\
As stable quasicrystals are characterized by a high degree of order\cite{PhysicalProperties.1999.Stadnik}, we choose the spins to be at sites with a fixed coordination number $z$. 

An example for the susceptibility $\chi_{l,m}$ is shown in Fig.~\ref{fig:tilings} for each of the two tilings. In general, the susceptibility strongly depends  on $E_\mathrm{F}$ and on the local environment, and interactions of similar strength can be found whenever  the same local pattern reoccurs in the tiling. Since any local pattern of linear dimension $L$ is repeated in a distance $\mathcal{O}(L)$ for each quasiperiodic tiling, \cite{SciAm.1977.Gardner} the strongest bonds form clusters that have a fixed form and are repeated quasiperiodically.

In our previous work (Ref.~\onlinecite{EPL.2015.Thiem}) we studied the distribution of $\chi_{l,m}$ in detail for the Ammann Beenker tiling. We observe the same qualitative behaviour also for the Penrose tiling and we briefly summarize the main findings here before focussing on results from Monte Carlo simulations. In general, the susceptibility $\chi_{l,m}$ is proportional to the local DOS at $E_\mathrm{F}$ which is known to vary strongly with the local environment \cite{JPhysFrance.1994.Jagannathan}. This also means that interactions are smaller when $E_\mathrm{F}$ lies in a pseudogap which is a typical feature of quasicrystals. 
We find oscillations between ferromagnetic and antiferromagnetic interactions as a function of site separation $r$, measured along the shortest bond path. This behaviour differs significantly from that in periodic systems with a spherical Fermi surface, where interactions oscillate within a power-law envelope that varies as $r^{-d}$. Although in some cases the average $\langle \chi_{l,m} \rangle$ is reasonably well described by this power law, individual interactions in quasiperiodic tilings can be considerably larger, and we usually observe a wide range of interaction strengths at each $r$. This results in a quasi-random contribution to the magnetic interactions due to the many local environments in a quasicrystal. 

\section{Monte Carlo Simulation}
\label{sec:pt}

We use Monte Carlo simulations to study the statistical mechanics of the spin Hamiltonian $H_\mathrm{RKKY}$. We apply parallel tempering which simulates multiple copies of the system at different temperatures $T_i$, so reducing correlation times. \cite{PhysRevLett.1986.Swendsen, EurPhysLett.1992.Marinari} We use a geometric distribution of the temperatures at low temperatures with $T_m = c^m \, T_\mathrm{min}$ and a uniform spacing with $T_m = T_{m-1} + c_2$ above the phase transition. For each replica one Monte Carlo Sweep consists of $N$ single-spin flip attempts and one replica-swap attempt, where $N$ is the number of spins in the system. The number of replicas (typically 60-100) is chosen to have an acceptance ratio of at least 30\% for the replica swaps.

We compute a selection of observables to study the phase transition and the ground state of this system. This includes the energy $E =  \sum_{l,m} \chi_{l,m} \sigma_l \sigma_m $ and the magnetization  $M  =  \sum_{l=1}^{N} \sigma_l $. Thermal averages are computed over $5\times 10^5$ Monte Carlo Sweeps. The heat capacity per spin is $C = \frac{1}{NT^2} \left[ \langle E^2\rangle - \langle E\rangle^2 \right]$ and the susceptibility per spin is $\chi = \frac{1}{NT}\left[ \langle M^2\rangle - \langle |M|\rangle^2 \right]$. The former provides a signal for phase transitions. While for conventional antiferromagnets the staggered magnetization is a useful tool to detect long-range order, we do not know the structure of the ground state a priori. To look for long-range magnetic order in an unbiased way, we first search for the ground state configuration $\{\xi_l\}$ and then compute the corresponding order parameter $M_\mathrm{gs} = \sum_{l=1}^N \xi_l \sigma_l$. We can use $M_\mathrm{gs}$ to define the order parameter susceptibility $\chi_\mathrm{op} = \frac{1}{NT}\left[ \langle M_\mathrm{gs}^2\rangle - \langle |M_\mathrm{gs}|\rangle^2 \right]$.

Further, we consider some quantities commonly used to study spin glasses which turn out to be sensitive to different ordering patterns. This includes the overlap $q = \frac{1}{N} \sum_{l=1}^{N}  \sigma_l^1 \sigma_l^2 $ and its distribution $P(q)$ obtained from the simulations of two independent replicas with the same interactions and temperature. We find that the Binder cumulant $B_\mathrm{SG} = \frac{1}{2} (3 - \langle q^4\rangle / \langle q^2 \rangle^2 )$ is a good tool to identify order, whether it is (anti)ferromagnetic or spin-glass-like. This becomes clear by looking at the limiting cases: at high temperatures the spins are uncorrelated and we obtain $q\sim N^{-1/2}$ and $B_\mathrm{SG}=0$; in contrast, if there is only a pair of ground states related by a global spin inversion at low temperatures, we obtain $q=\pm 1$ and $B_\mathrm{SG}=1$.

\subsection{Phase Transition}

We use Monte Carlo simulations for different parameter sets for the Ammann Beenker tiling and the Penrose tiling obtained by varying $E_\mathrm{F}$ and choices $z$ for the coordination number of the magnetic sites. Our results are briefly summarized in Table \ref{tab:overview}. We were able to classify the systems by the structure of their ground state into two main classes: 
The first class is characterized by ground states with large ferromagnetic regions and we denote it as class FMR. The second class (class AF) shows ground states with antiferromagnetic correlation. We can further subdivide this class with respect to the number of ground state. While class AF1 has only a single pair of ground states related by time-reversal symmetry, the class AF2 is characterized by multiple low-energy states. This leads to distinct differences in the statistical mechanics of the systems as outlined below. In the following we discuss the classes in detail.

Due to the computational complexity of our simulations, we can only obtain simulations results for very few successive approximants for each system (see also Sec. \ref{sec:fss}). For these successive approximants we find the same qualitative behaviour provided they are big enough (about 300 spins) to capture the physics of the quasiperiodic system. 

\begin{table}
\caption{Classification of low-temperature states for different Fermi energies $E_\mathrm{F}$ and spins at sites with coordination numbers $z$ for the Ammann Beenker tiling and the Penrose tiling. (We did not study the Penrose tiling with $z=4$ because this system has a very inhomogeneous distribution of spins.) Abbreviations: FMR - ferromagnetic regions, AF1 - antiferromagnetic with a single ground state, AF2 - antiferromagnetic with multiple ground states.}
\label{tab:overview}
\begin{tabular}{ p{1.8cm} p{1cm} p{1.5cm} p{1.5cm} p{1.5cm} }\hline\hline
system	& $E_\mathrm{F}$ &  $z=3$ & $z=4$ & $z=5$ \\\hline
Penrose & -4    & FMR& - & FMR \\
			& -3.5 & AF1       & - & AF2 \\
			& 0      & AF1 & - & AF1  \\
			& 0.07 & AF2      & - & AF1  \\
			& 2.33 & AF2 & - & AF1 \\
			& 2.79 & AF1 & - & AF2 \\ \hline
Ammann	  & -4    & FMR & FMR     &  AF1  \\
Beenker   	& -3.5 & AF2 & AF2 & AF1 \\
				& 0		 & AF1 & FMR & AF1 \\
				& 0.7	& AF2 & AF2 & AF1 \\
				& 1.12 & AF1 & AF2 & AF2  \\
				& 1.3   & AF2 & AF2 &  AF2 \\
				& 1.95 & AF2 & AF1       & AF2 \\
				& 2.7   & AF2 & AF1 &  AF2 \\ \hline\hline
\end{tabular}
\end{table}

\subsubsection{FMR class}

Systems with large ferromagnetic regions mainly occurs for Fermi energies near the band edges. In these cases, the dominant contributions to $\chi_{l,m}$ are from electronic wave functions which have a slowly varying envelope and components on the two sublattices  of the tiling with either the same sign (upper band edge) or opposite signs (lower band edge). Both cases lead to dominant ferromagnetic interactions between the spins with very weak frustration. 

\begin{figure}
    \includegraphics[width=\columnwidth]{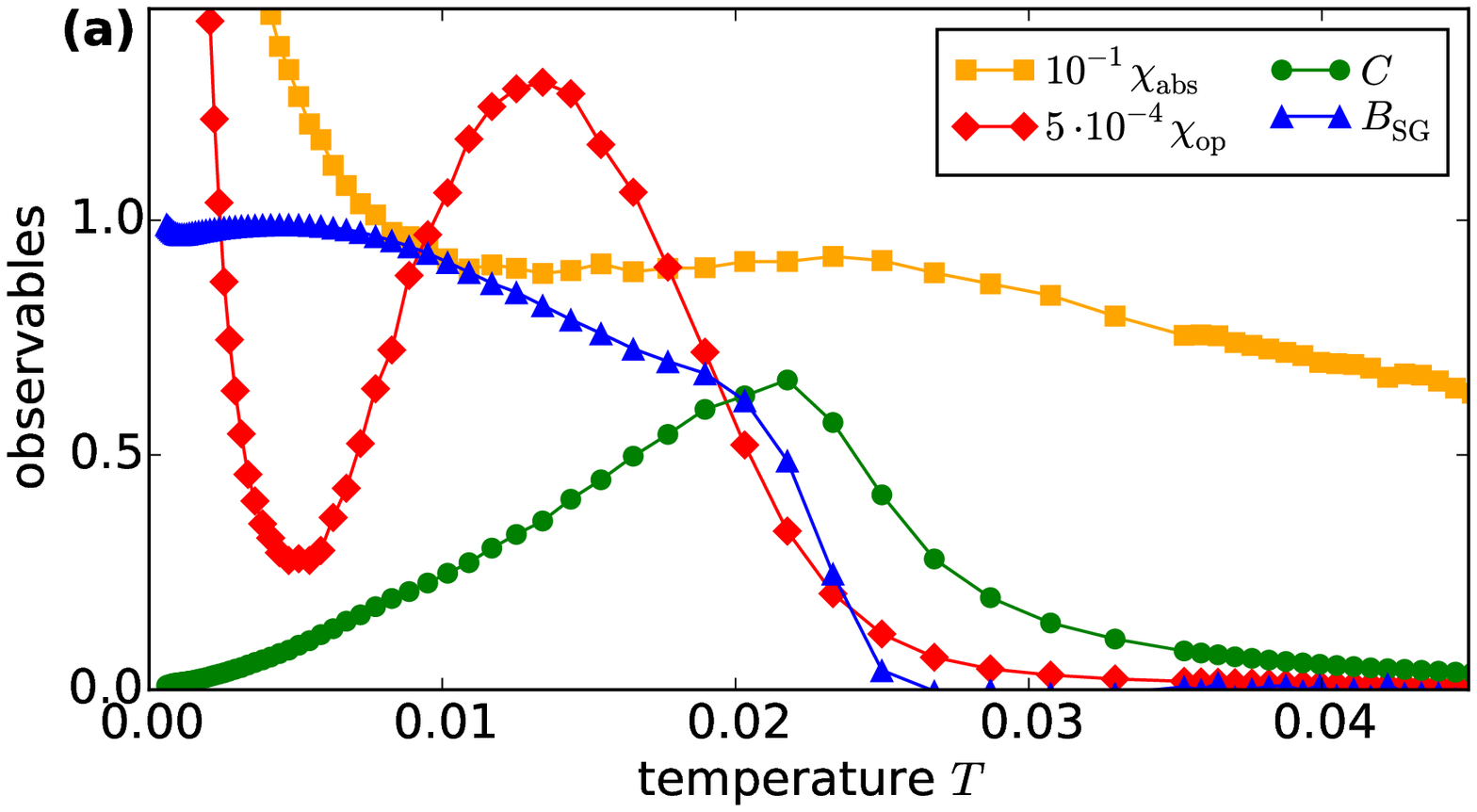}
    \includegraphics[width=\columnwidth]{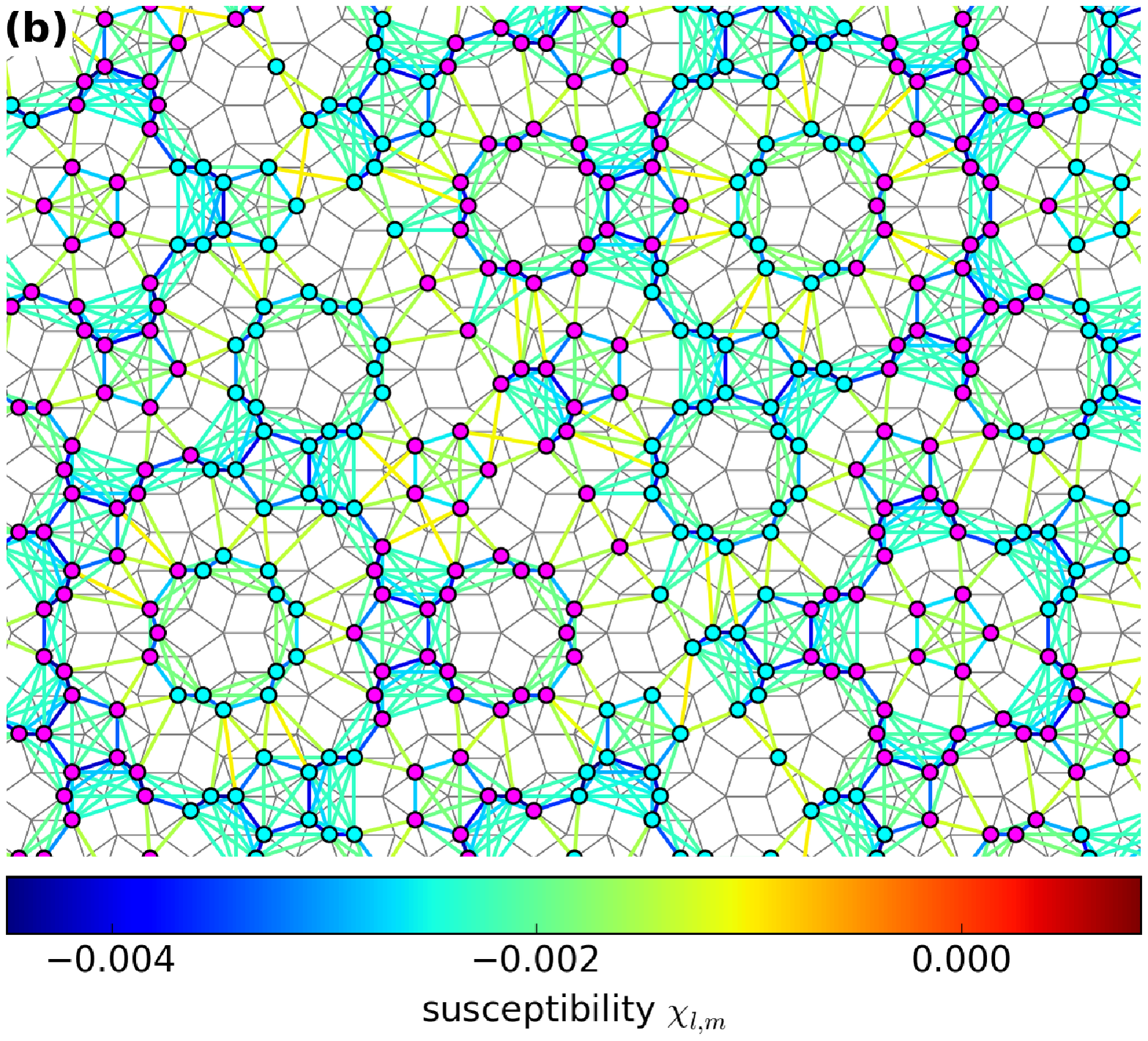}
     \includegraphics[width=\columnwidth]{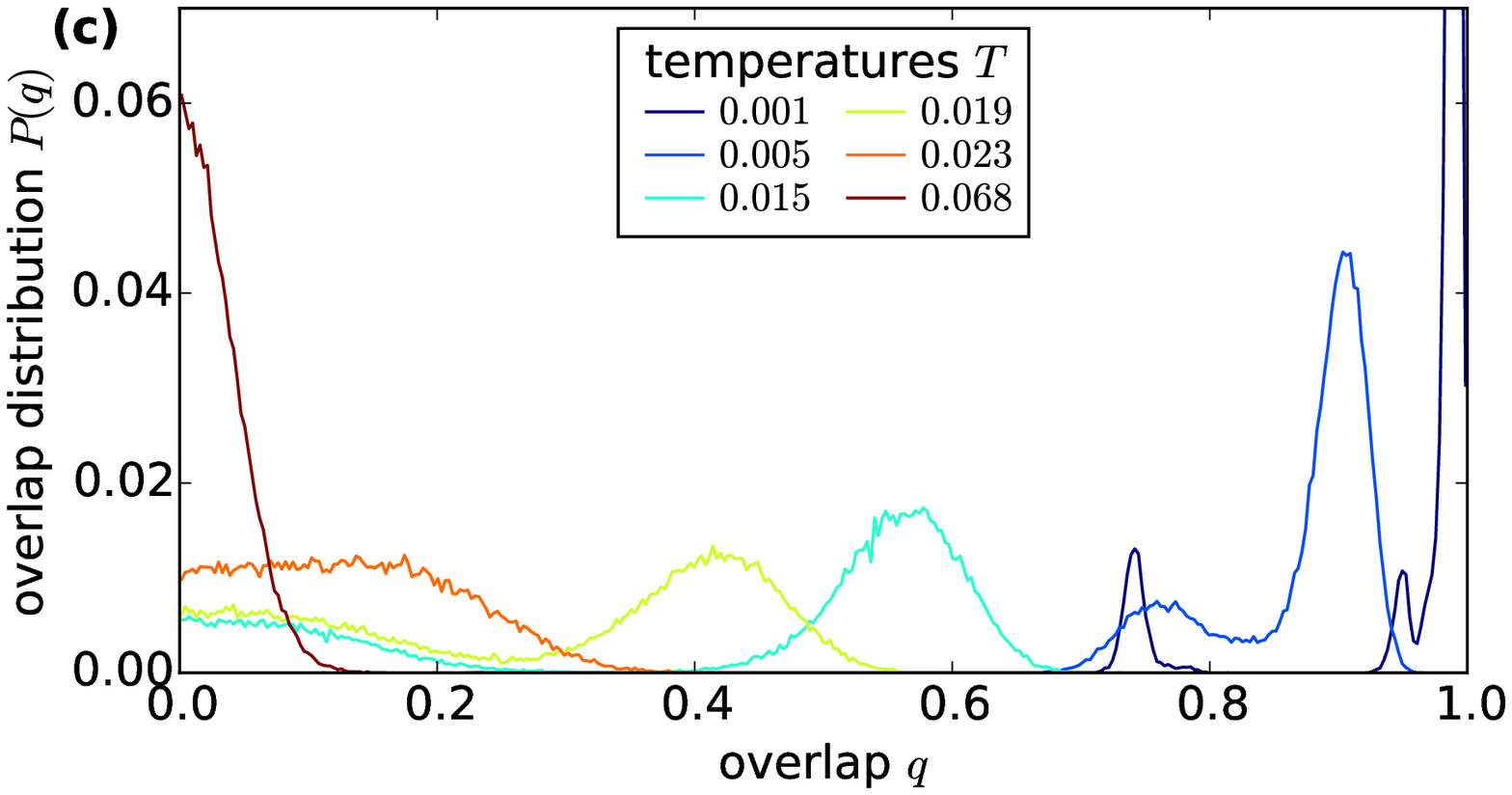}    
\caption{Simulation results for the Penrose tiling with $z=3$ and $E_\mathrm{F} = -4$ (class FMR): (a) different observables and (b) lowest energy spin configuration (red and blue circles for up and down spins) and (c) overlap distribution $P(|q|)$ at different temperatures.} 
\label{fig:pt-penrose-fm}
\end{figure}

A typical example is the Penrose tiling illustrated in Fig.~\ref{fig:pt-penrose-fm} for the Fermi energy $E_\mathrm{F} = -4$ with $N = 693$ Ising spins at sites with coordination number $z=3$. The ground state shows large ferromagnetic regions and the weak long-range antiferromagnetic bonds lead in this case to the formation of large oppositely-aligned regions. Often the spin patterns form rather elongated regions in order to avoid domain walls along strong ferromagnetic bonds. According to the results in Fig.~\ref{fig:pt-penrose-fm}a, the system shows a phase transition at about $T_\mathrm{f} \approx 0.02$ from a paramagnetic state at high temperatures to an ordered state at low temperatures, as suggested by the broad peaks in the heat capacity $C$ and the susceptibility $\chi$ as well as an increase of the Binder cumulant. At very low temperatures the susceptibility $\chi$ and the order parameter susceptibility $\chi_\mathrm{op}$ show a $1/T$-divergence due to the fluctuations of small clusters or single spins. This is discussed in more detail in Sec.~\ref{sec:fluctuations}.

\subsubsection{AF classes}

Most of the studied systems belong to the AF class for which we find a phase transition to low-temperature states with antiferromagnetic correlations. In the following we focus on this class, and in Sec.~\ref{sec:structureAFM} we show that the low-temperature state has quasiperiodic N\'eel order. To illustrate this transition, we first present simulation results for both sub-classes, AF1 and AF2.

\begin{figure}
    \includegraphics[width=\columnwidth]{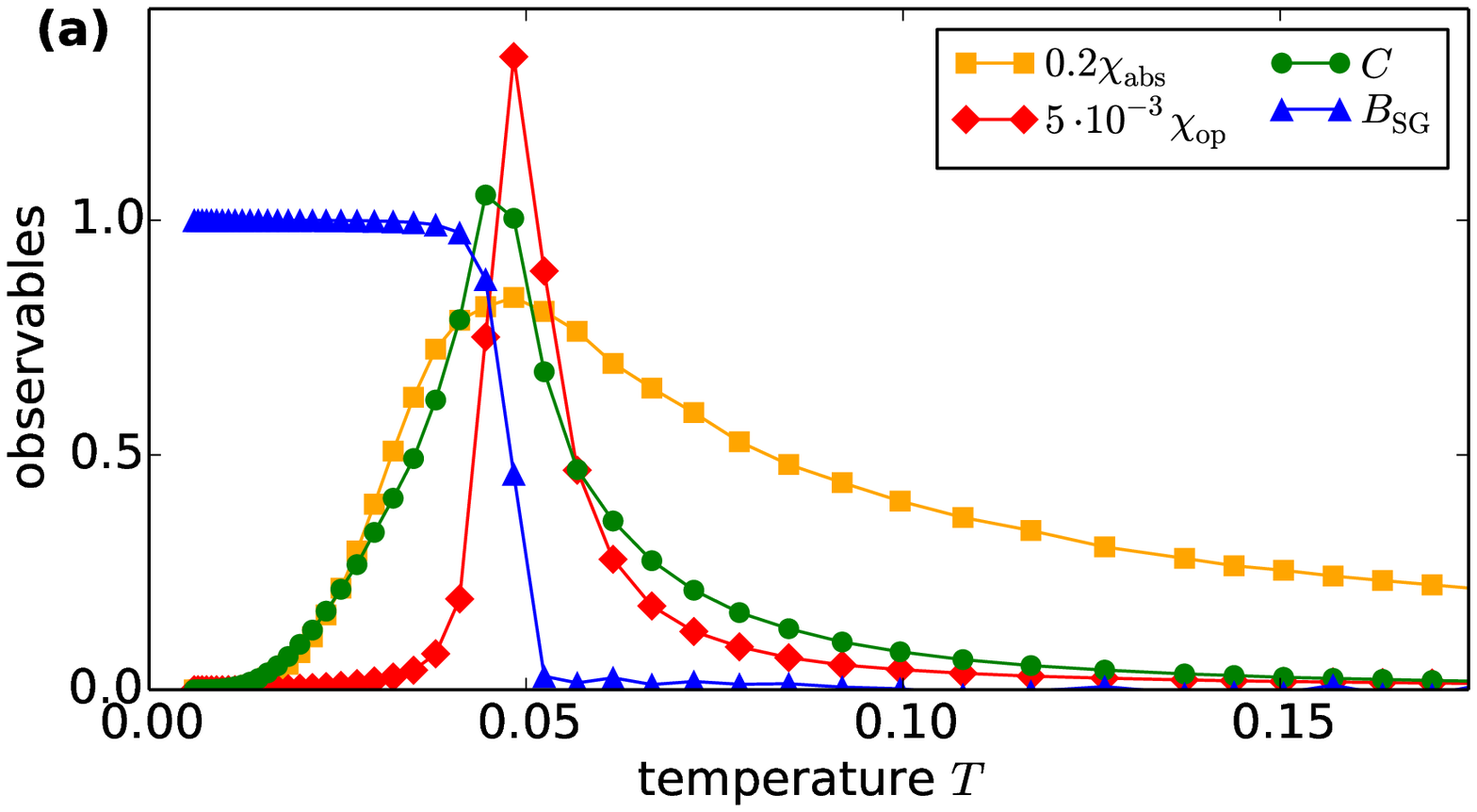}
    \includegraphics[width=\columnwidth]{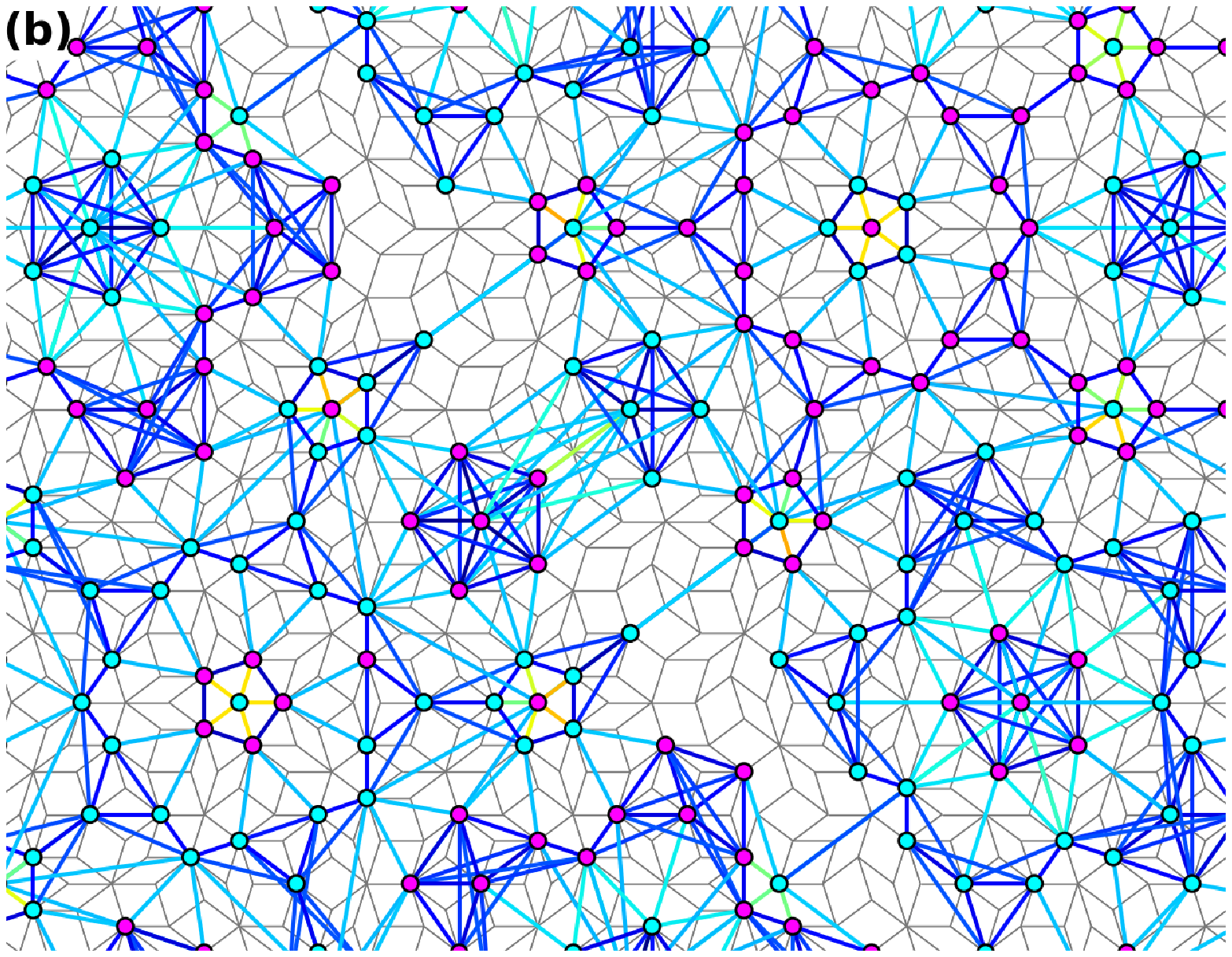}
     \includegraphics[width=\columnwidth]{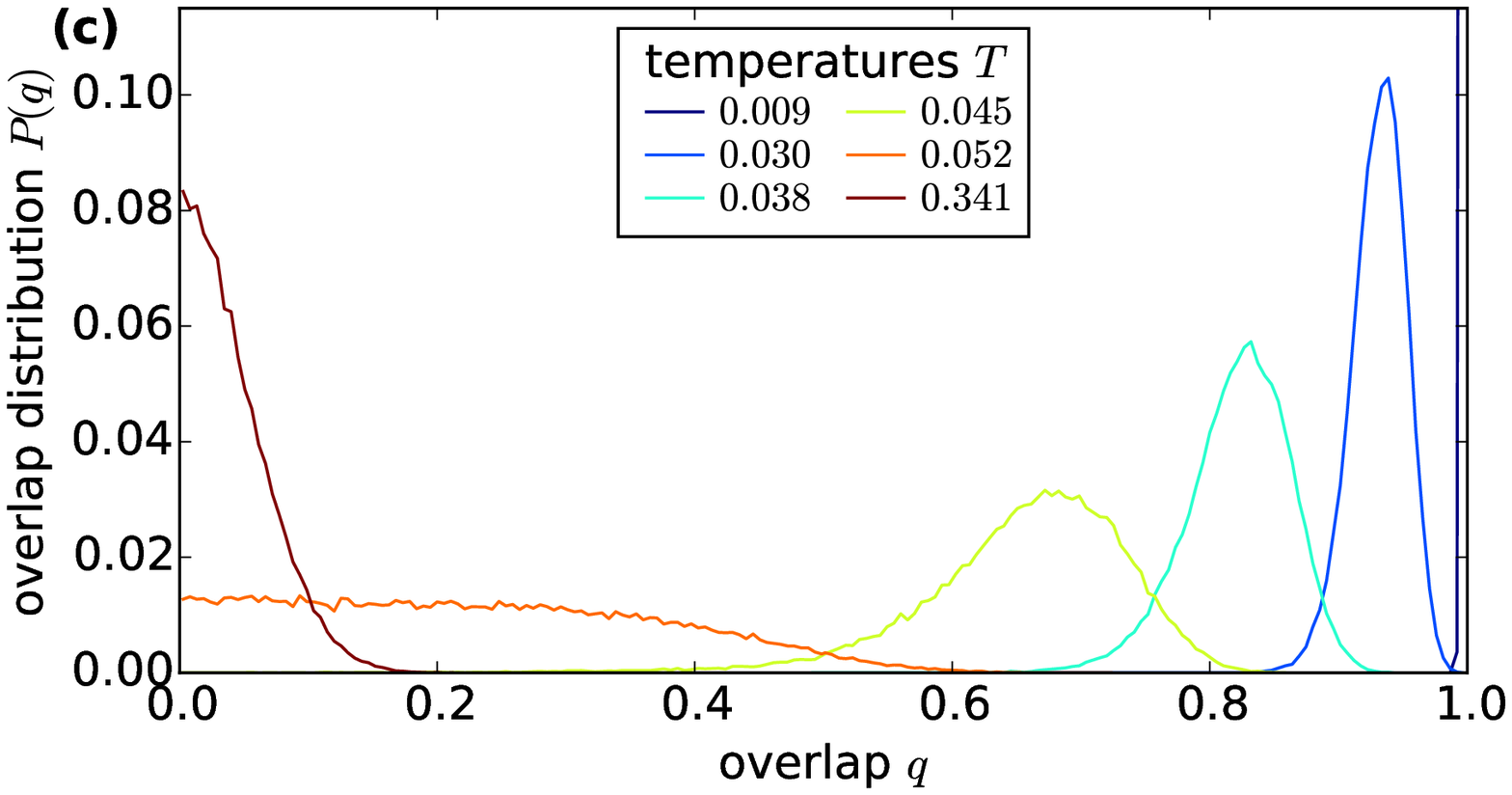}    
\caption{Simulation results for the Penrose tiling with $z=5$ and $E_\mathrm{F} = 0.07$ (class AF1): (a) different observables and (b) lowest energy spin configuration (red and blue circles for up and down spins) with strongest bonds according to Fig.~\ref{fig:tilings}a and (c) overlap distribution $P(|q|)$ at different temperatures.} 
\label{fig:pt-penrose}
\end{figure}

A typical example for the AF1 class is shown in Fig.~\ref{fig:pt-penrose}. It corresponds to the Penrose tiling in Fig.~\ref{fig:tilings}a with $N = 375$ Ising spins at sites with coordination number $z=5$ and taking $E_\mathrm{F} = 0.07$. For this system, the heat capacity $C$ has a peak near $T_\mathrm{f} \approx 0.05$, and the Binder cumulant $B_\mathrm{SG}$ approaches 1 below this temperature. This suggests that the system is paramagnetic for $T > T_\mathrm{f} $, and that at $T_\mathrm{f} $ a macroscopic fraction of  spins lock together into a state with long-range rigidity. There is a large peak in the order parameter susceptibility  $\chi_\mathrm{op}$ at $T_\mathrm{f}$ associated with the spontaneous breaking of the global Ising symmetry of the spin model when spins order into a low-temperature phase with nearly zero magnetization $M$ but non-zero $M_{\rm gs}$. In our previous work (see Ref.~\onlinecite{EPL.2015.Thiem}) we found the same qualitative features for the Ammann Beenker tiling with $E_\mathrm{F} = 1.95$ and $z=4$.

\begin{figure}
    \includegraphics[width=\columnwidth]{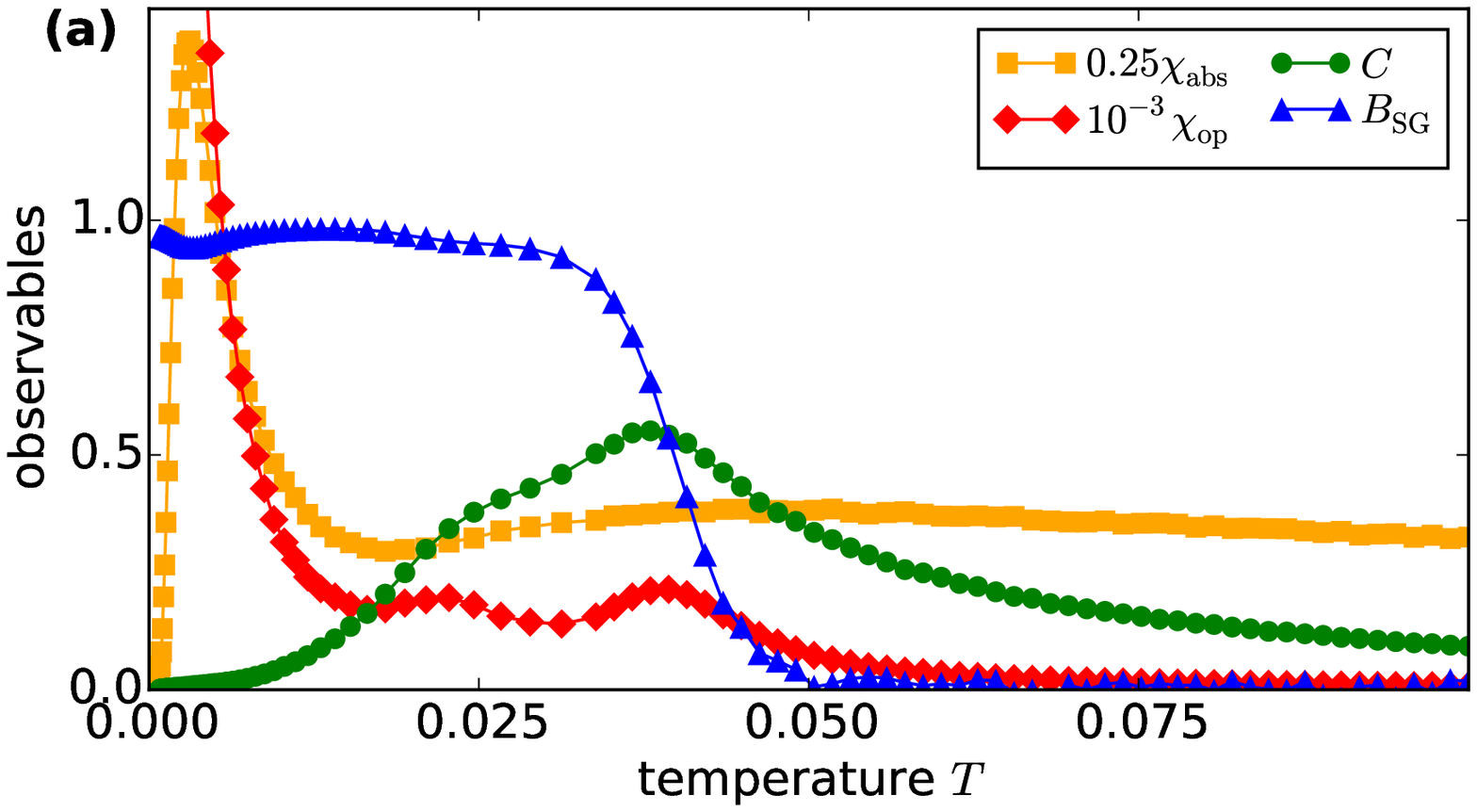}
    \includegraphics[width=\columnwidth]{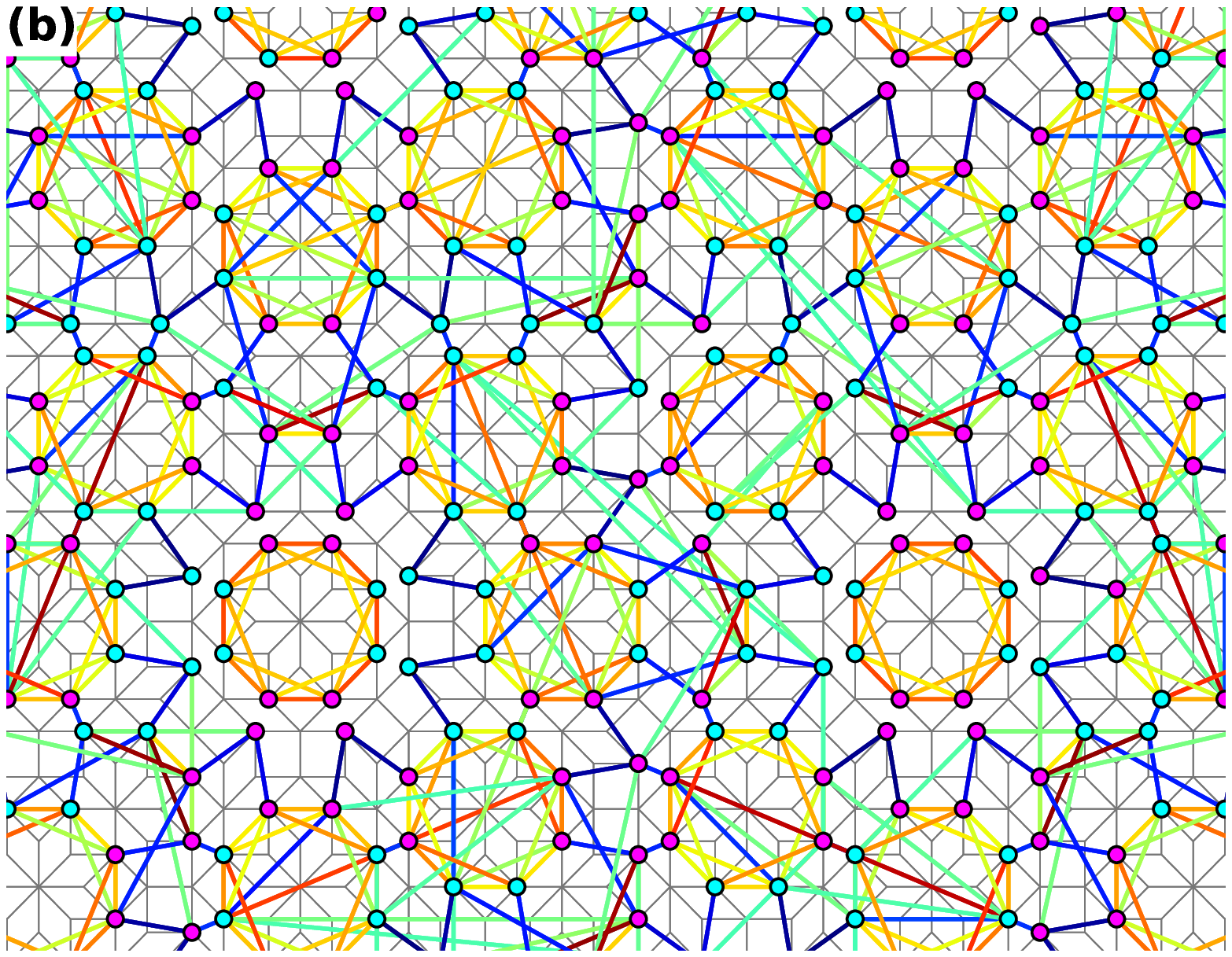}
    \includegraphics[width=\columnwidth]{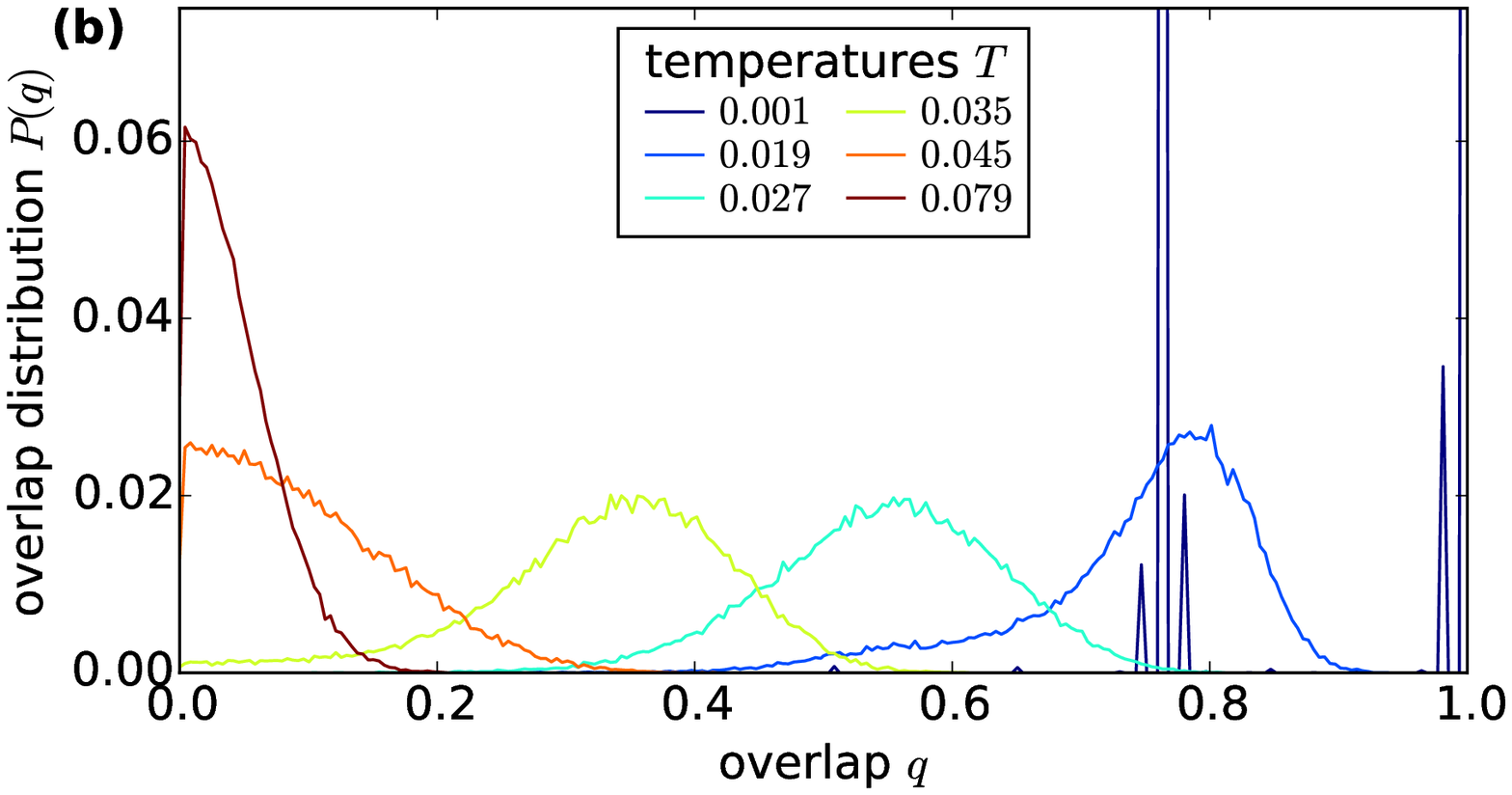}
\caption{Simulation results for the Ammann Beenker tiling with $z=4$ and $E_\mathrm{F} = 1.12$ (class AF2): (a) different observables and (b) lowest energy spin configuration (red and blue circles for up and down spins) with strongest bonds according to Fig.~\ref{fig:tilings}b and (c) overlap distribution $P(|q|)$ at different temperatures.}
\label{fig:pt-octagonal}
\end{figure}

In contrast, behaviours in the AF2 class are more complex. An example is the Ammann Beenker tiling with $N = 478$ spins at sites with $z=4$ and $E_\mathrm{F} = 1.12$ in Fig.~\ref{fig:tilings}b. The temperature behaviour of the observables and ground state are shown in Fig.~\ref{fig:pt-octagonal}. Clear indications for a phase transitions come from the Binder cumulant $B_\mathrm{SG}$ which increases from 0 to values above 0.95 below  $T_\mathrm{f} \approx 0.04$. Again this is due to the transition from a paramagnetic state for $T > T_\mathrm{f}$ to a state with broken Ising symmetry with antiferromagnetic correlations below $T_\mathrm{f}$. The heat capacity $C$ and the susceptibility also show very broad maxima near $T_\mathrm{f}$. 

To distinguish between the two AF subclasses, we show the overlap distribution $P(q)$ in Figs.~\ref{fig:pt-penrose}c and \ref{fig:pt-octagonal}c. For both systems $P(q)$ is centred at $q=0$ for $T>T_\mathrm{f}$ but develops one or more pairs of peaks at $\pm q_i(T)$ for $T<T_\mathrm{f}$. If we observe a single pair of peaks $q_0 \to \pm 1$ at very low temperatures, the spins freeze in a particular ground state. The system in Fig.~\ref{fig:pt-penrose}c is an example of this behaviour. In contrast, the overlap distribution in Fig.~\ref{fig:pt-octagonal}c shows 4 peaks at $q_0=\pm 1$ and $q\approx \pm 0.75$ for $T \to 0$, indicating that there are two competing pairs of low-temperature states. We use the standard deviation of the overlap distribution $\sigma(P(|q|))$ at low temperatures to distinguish the two classes. We find that $\sigma = 0.02$ is a good threshold value to distinguish the AF2 class with multiple low-temperature states ($\sigma > 0.02$) from the AF1 class with a single ground state ($\sigma  < 0.02$). 

For the AF2 class, signatures of the multiple ground states are also clearly visible in the order parameter susceptibility $\chi_\mathrm{op}$ which diverges at low-temperature due to a Curie-like contribution originating from the fluctuations of the system between the different ground states (see Fig.~\ref{fig:pt-octagonal}). Also the susceptibility $\chi$ of this system has an additional rather sharp peak at $T_1 \approx 0.005$. In Sec.~\ref{sec:fluctuations} we show that this peak is caused by fluctuations of loosely-coupled spins.

Note that the overlap distributions we obtain are strikingly different from those observed in spin glass simulations using the three-dimensional Edwards Anderson model or the Sherrington-Kirkpatrick model \cite{PhysRevLett.2012.Yucesoy} for which $P(q)$ has much more weight at smaller $q$ below $T_\mathrm{f}$. 

Although we can access only a few approximants numerically, we expect that the number of low-temperature states in AF stays finite in the thermodynamic limit. Evidence for this comes from the finite size scaling results in Sec.~\ref{sec:fss} that show an excellent agreement of the different approximants according to relevant scaling equations.

\subsection{Structure of Ground State}
\label{sec:structureAFM}

A key question is what type of order characterises the low-temperature phase. Short-distance correlations are readily apparent in the ground state spin configuration. For example, the Penrose tiling in Fig.~\ref{fig:pt-penrose}b contains many pentagons with a ferromagnetic spin configuration forming on the 5-fold symmetric stars of the tiling. Long-range order results from a tendency of neighbouring clusters to anti-align due to antiferromagnetic bonds between the clusters. For the Ammann Beenker tiling in Fig.~\ref{fig:pt-octagonal}b two distinct local patterns can be identified: octagons with an antiferromagnetic spin configuration form on the 8-fold star patterns, and small V-shaped clusters form along the strong ferromagnetic bonds. Again long-range order is caused by a tendency of neighbouring clusters to anti-align. 

In both cases the nature of this order is revealed to be quasiperiodic by the magnetic structure factor $M(\mathbf{k}) = \left|\sum_l e^{2\pi \mathrm{i}\cdot \mathbf{k} \mathbf{r}_l} \xi_l \right|^2$ computed from the spin configuration $\{\xi_i\}$ of the ground state. In contrast to periodic systems, quasicrystals do not posses a Brillouin zone and understanding the diffraction pattern is usually challenging. However, it is possible to derive the nuclear structure factor of quasiperiodic tilings theoretically, and one can show that the quasiperiodicity leads to Bragg peaks in a pattern that forms a reciprocal-space quasiperiodic tiling. \cite{AperiodicOrder.2013.Baake} The corresponding length scale is $4\pi \tau /(5b)$ for the reciprocal-space Penrose tiling\cite{ActaCrys.2010.Kozakowski} and $1/(2b)$ for the reciprocal-space Ammann Beenker tiling\cite{ActaCrysA.1988.Wang, CrysQuasicrystals.Steurer} in units of the real-space bond length $b$ and the golden mean $\tau = (1+\sqrt{5})/2$. When only a subset of sites of the tiling is chosen, the nuclear structure factor is rather robust and there are often only minor changes in the amplitudes of the Bragg peaks. \cite{JPhysFrance.1989.Sire2}
 
\begin{figure}
    \centering
    \includegraphics[width=0.97\columnwidth]{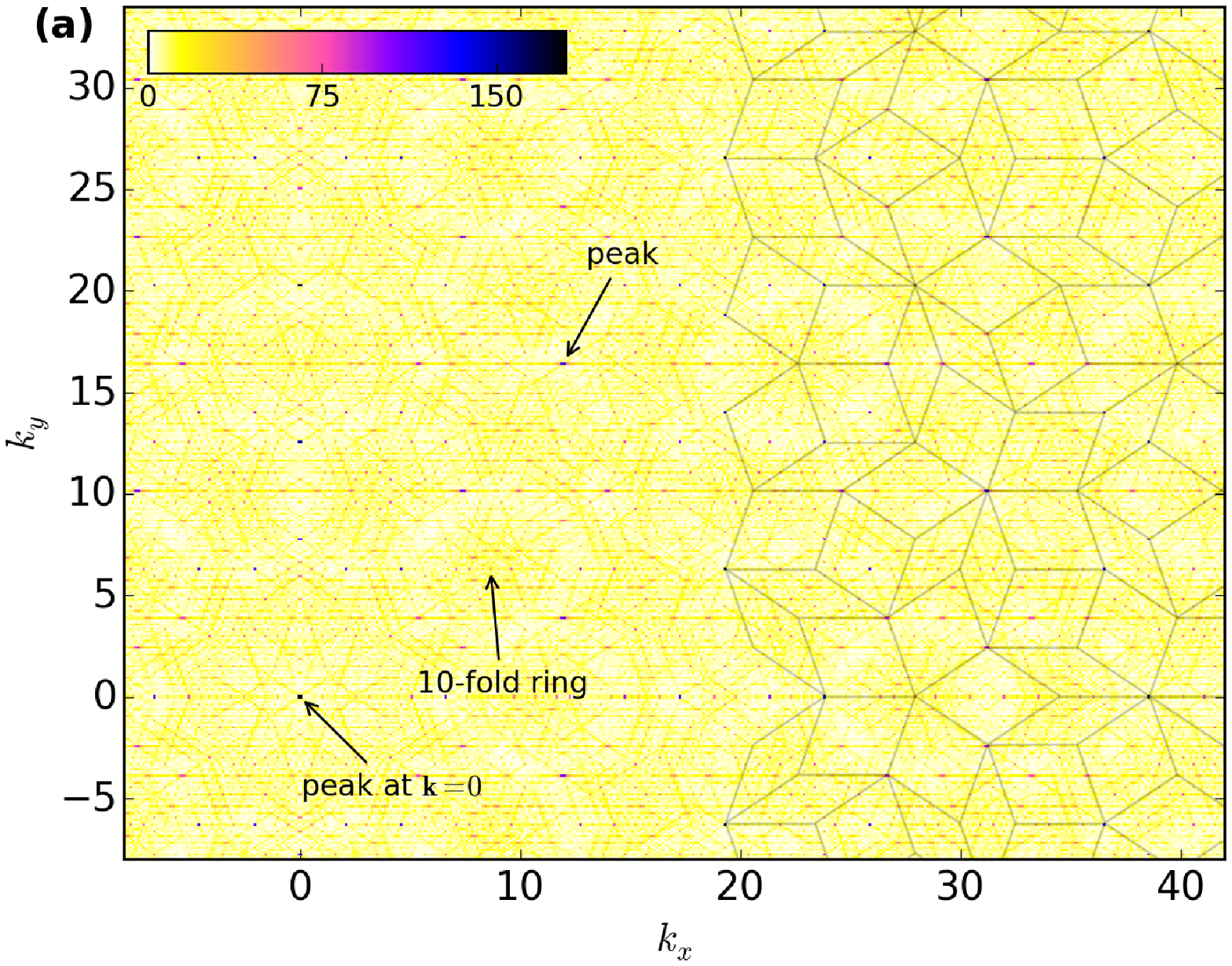}
    \includegraphics[width=0.97\columnwidth]{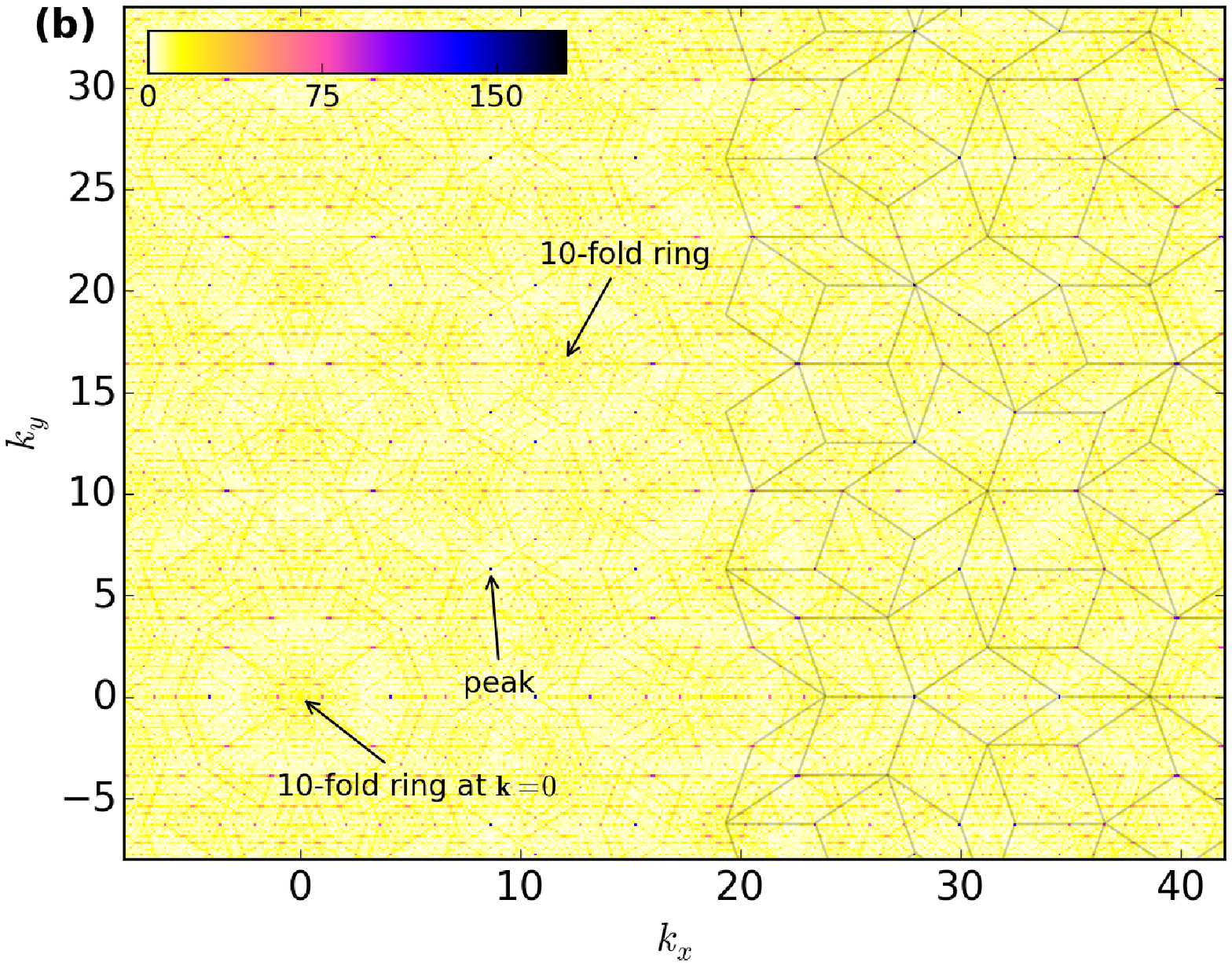}
    \includegraphics[width=0.97\columnwidth]{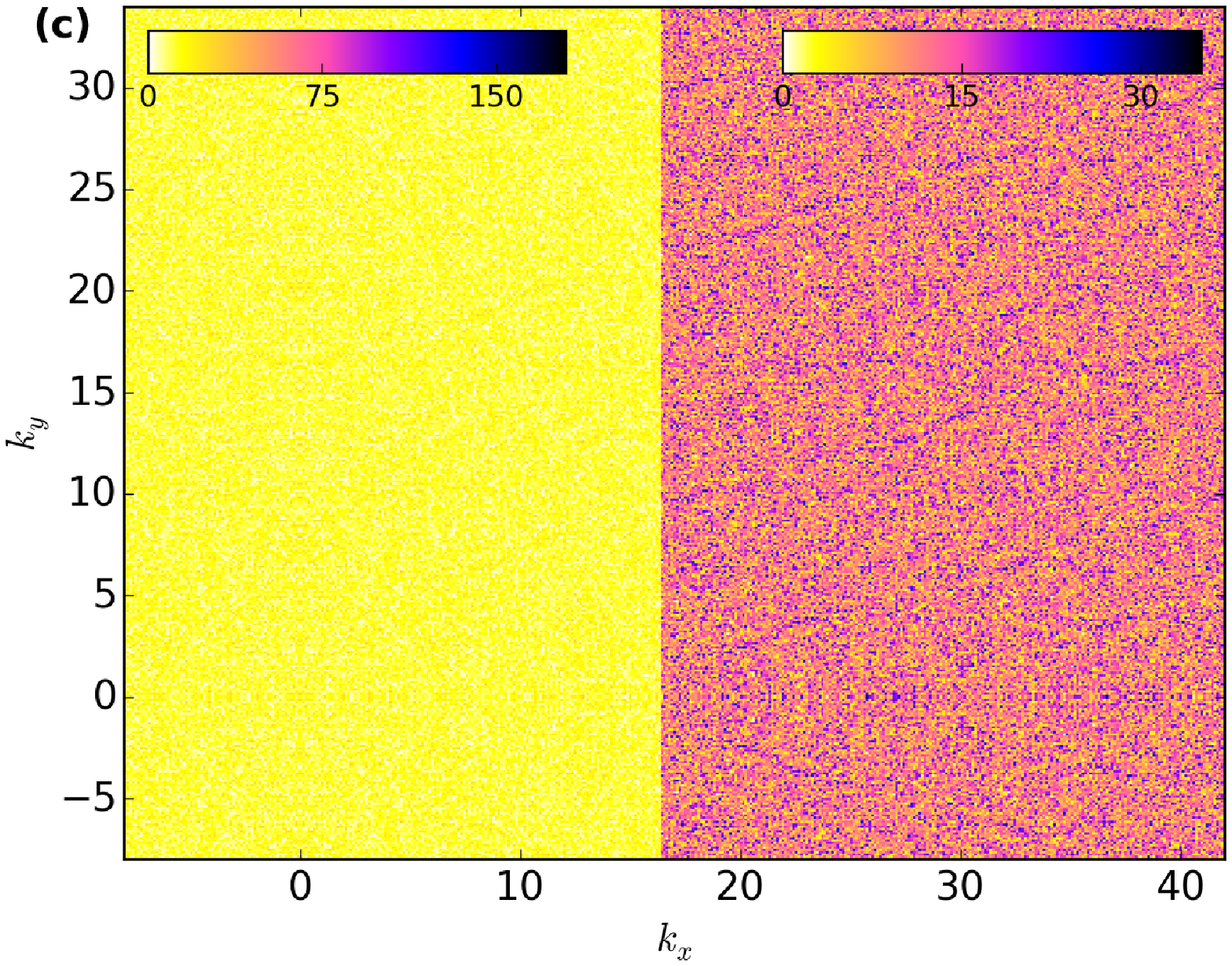}
\caption{Magnetic structure factor (scaled as $M(\mathbf{k})^\frac{3}{8}$ for better visibility) of the 6th approximant of the Penrose tiling with $N = 1003$ spins on sites with coordination number $z=5$: (a) ferromagnetic state, (b) lowest energy spin configuration for $E_\mathrm{F} = 0.07$, and (c) random spin configuration. In the first two cases a Penrose tiling with the expected length scale $4\pi \tau /(5b) \approx 4.07$ for $b=1$ is overlaid on the right half of the image to visualize the quasiperiodic structure. Note that the peaks and 10-fold rings (marked by arrows) are exchanged between (a) and (b).}
\label{fig:ft-penrose}
\end{figure}

Numerical results for a ferromagnetic state (all $\xi_i = 1$) on the Penrose tiling with $z=5$ are shown in Fig.~\ref{fig:ft-penrose}a. This gives the nuclear structure factor, and the diffraction intensity $M(\mathbf{k})$ indeed shows a pattern of peaks which can be overlaid by a Penrose tiling with the expected length scale. All sites of the reciprocal-space Penrose tiling fall on one of two different patterns: a high-intensity peak or the centre of a ring with 10 medium-intensity peaks as highlighted in Fig.~\ref{fig:ft-penrose}a. The 10-fold rotational symmetry of the Penrose quasicrystal is also visible around the central Bragg peak at $\mathbf{k}=0$.

In the case of an antiferromagnetic ground state the correlations gives rise to selection rules that result in a shift of the reciprocal space indices. \cite{MatSciEng.2000.Lifshitz, PhysRevLett.2003.Wessel} For the example in Fig.~\ref{fig:ft-penrose}b, this means that all peaks split into several peaks along the principal directions of the tiling. This is clearly visible for the central Bragg peak which splits into 10 sizeable peaks along the principle directions of the Penrose tiling. All other high-intensity peaks for the ferromagnetic state in Fig.~\ref{fig:ft-penrose}a also split into 10-fold rings of medium-intensity peaks for the ground state with antiferromagnetic order. The opposite effect occurs for the 10-fold rings with medium-intensity peaks of the ferromagnetic state. While each of these peaks again splits into 10 peaks along the principal directions, constructive interference leads to the formation of a high-intensity peak in the centre of the ring in the antiferromagnetic ground state. The annotations in Fig.~\ref{fig:ft-penrose} highlight this transformation of the diffraction patterns (i.e.~a peak to a 10-fold ring and vice versa) between the ferromagnetic and antiferromagnetic spin configuration. The observed structure factors are also consistent for different approximants, i.e., the peaks occur at the same positions and become sharper for bigger approximants.

A possible concern is whether the quasiperiodic site locations by themselves might be sufficient to generate these features in the diffraction pattern, without magnetic order. To test this we have computed the magnetic structure factor for a random spin configuration on the tiling (Fig.~\ref{fig:ft-penrose}c): its amplitude fluctuations are Gaussian\cite{EPL.2015.Thiem} and it does not show any of the high intensity peaks that are present for the ground state and for a ferromagnetic one. This clearly indicates that the observed pattern is due to the long-range magnetic order and not due to the atomic order.

\subsection{Formation and Fluctuations of Spin Clusters}
\label{sec:fluctuations}

\begin{figure}
    \includegraphics[width=\columnwidth]{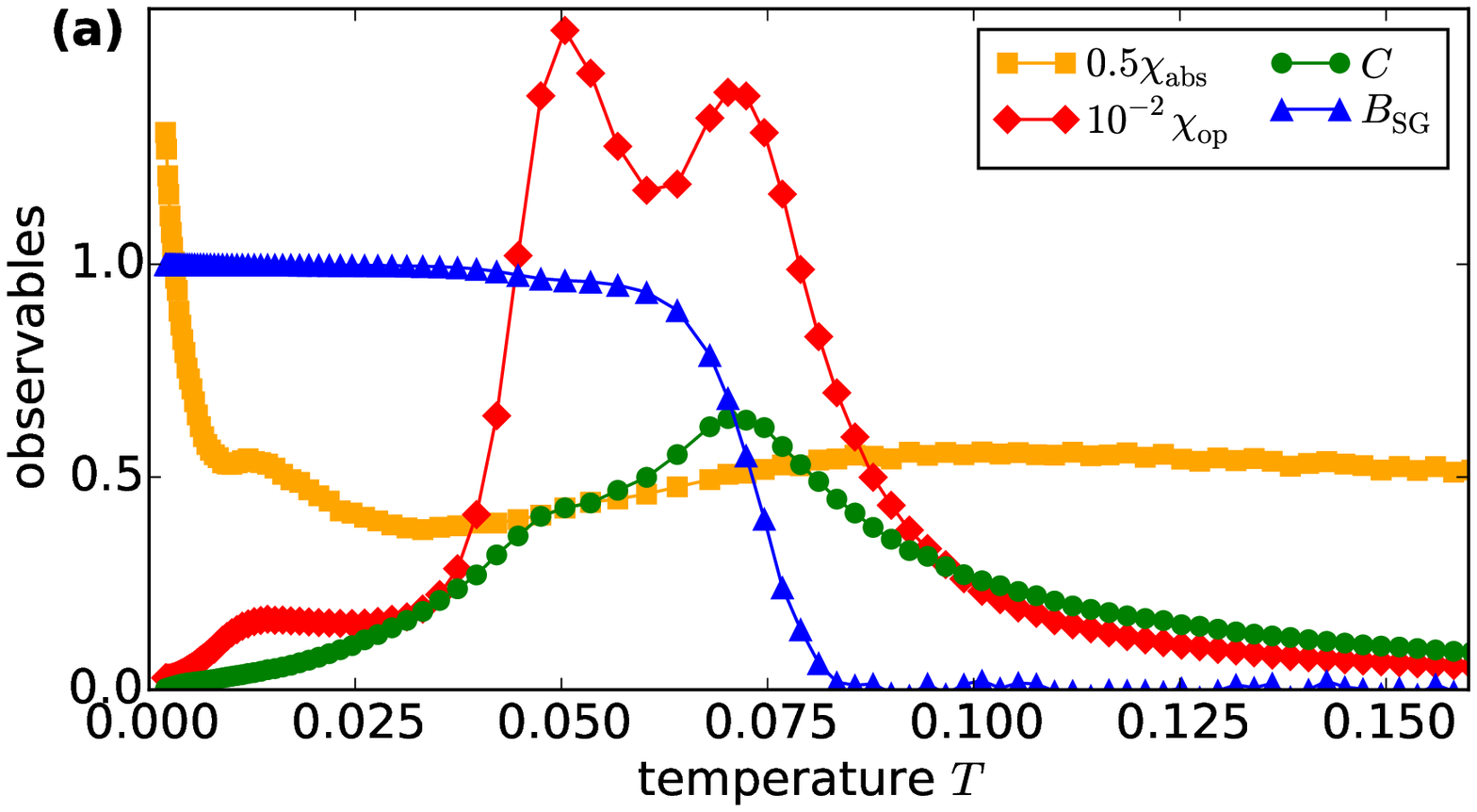}
    \includegraphics[width=\columnwidth]{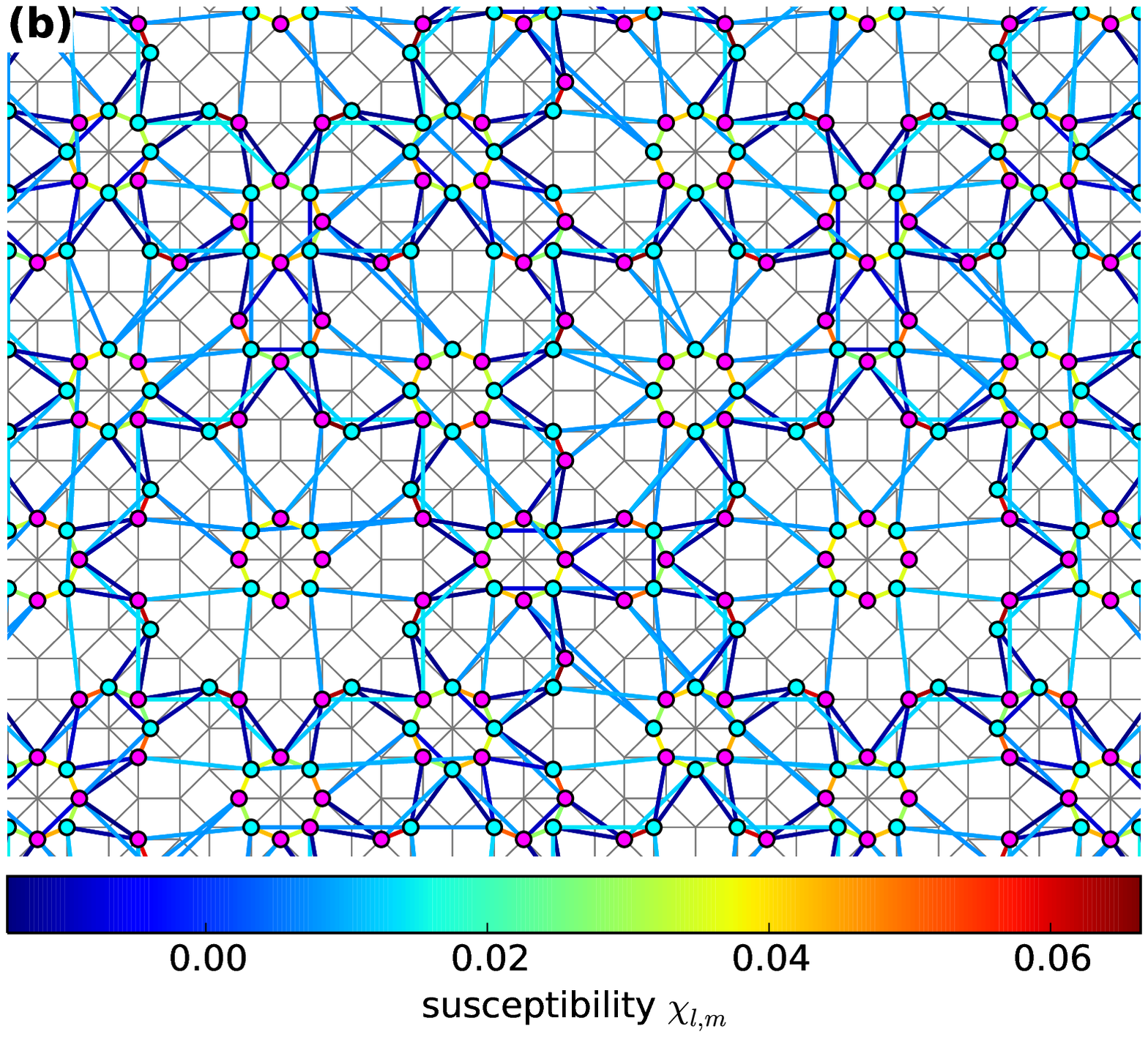}
    \includegraphics[width=\columnwidth]{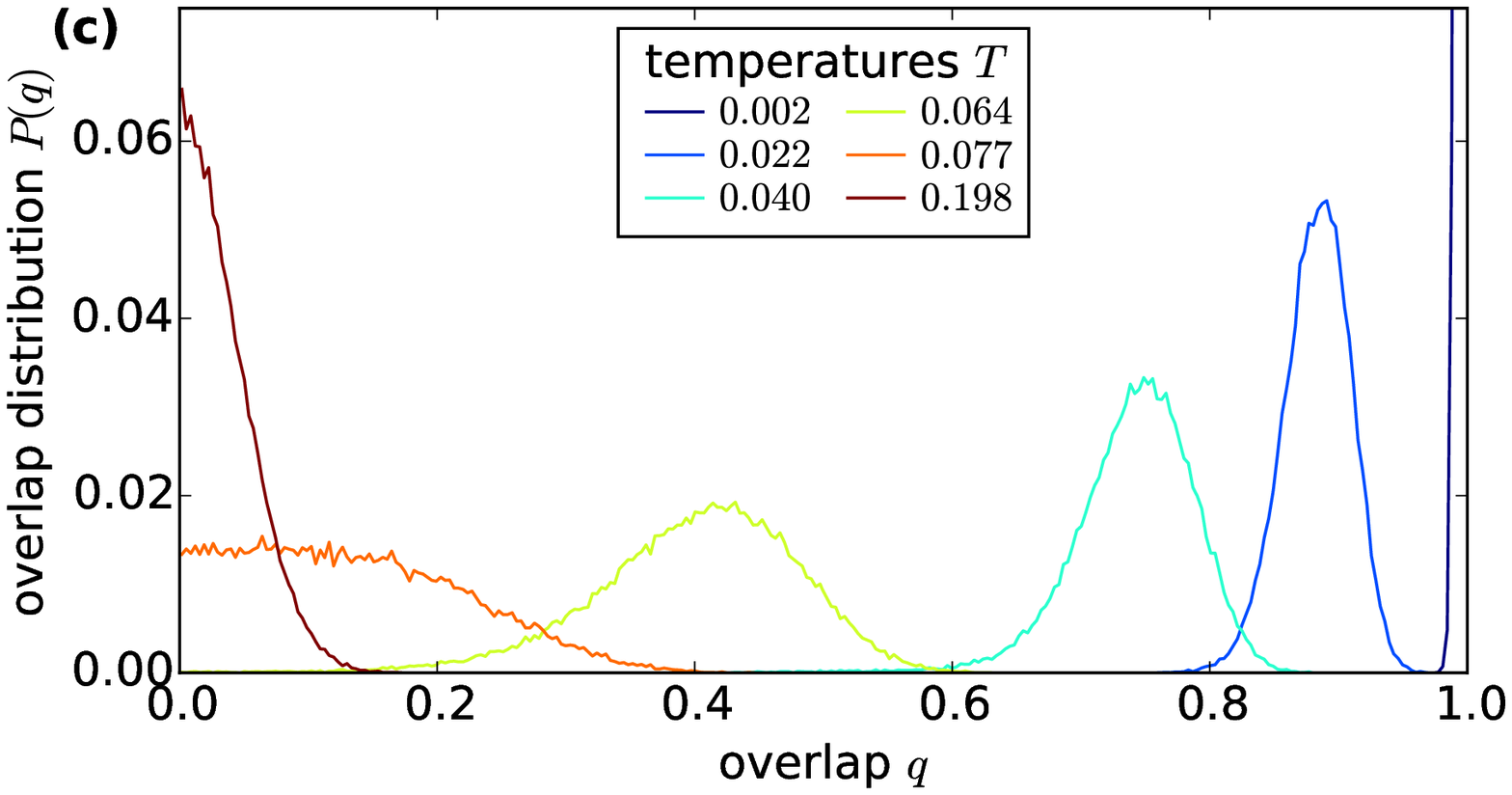}
\caption{Simulation results for the Ammann Beenker tiling with $z=3$ and $E_\mathrm{F} = 1.12$ (class AF1): (a) different observables, (b) lowest energy spin configuration (red and blue circles for up and down spins), and (c) overlap distribution.}
\label{fig:pt-octagonal2}
\end{figure}

\begin{figure*}
    \includegraphics[width = \textwidth]{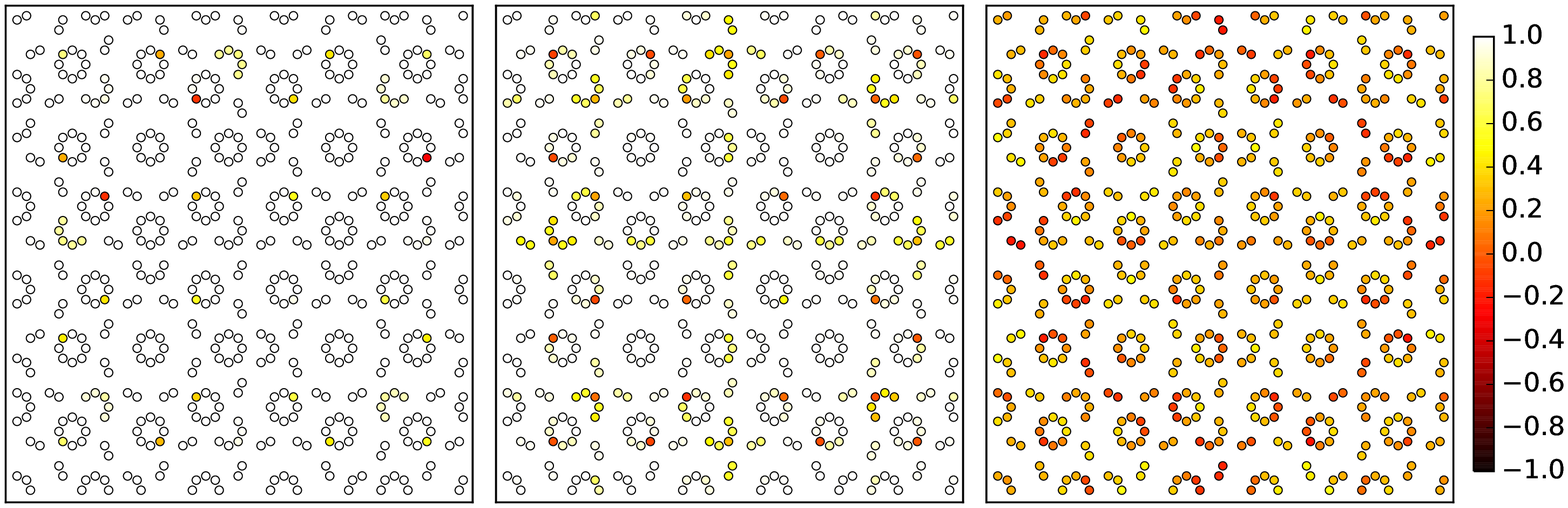}
\caption{Fluctuations $\langle m_\mathrm{gs}^i \rangle $ of spins for the system in Fig.~\ref{fig:pt-octagonal2} at different temperatures: (left) for $T = 0.01$ loose spins and semi-circles of spins are able to fluctuate leading to the peak in $\chi$; (middle) at $T = 0.031$ additional fluctuations occur, and (right) at the transition temperature  $T = 0.078 \lesssim T_\mathrm{f}$ fluctuations are present at all sites but $\langle m_\mathrm{gs}^i \rangle > 0$.} 
\label{fig:fluctuations}
\end{figure*}

Our results show some distinctive features in addition to those arising from ordering. In all systems considered, the RKKY interactions generate strongly coupled spin clusters with weaker inter-cluster couplings as illustrated in Fig.~\ref{fig:tilings}. For a given choice of magnetic site and $E_{\mathrm{F}}$ the most prominent clusters have a fixed form that is frequently repeated. According to Conway's theorem, the prominent clusters repeat quasiperiodically because any local pattern of linear dimension $L$ is repeated in a distance $\mathcal{O}(L)$ for each quasiperiodic tiling. \cite{SciAm.1977.Gardner} 

In addition, it is striking that in many systems some spins or small clusters are free to fluctuate at temperatures much below the ordering transition, yielding a Curie-like contribution to $\chi$ and/or $\chi_\mathrm{op}$. A typical example is the Ammann Beenker tiling with $z=3$ and $E_\mathrm{F} = 1.12$ shown in Fig.~\ref{fig:pt-octagonal2}. The system belongs to the class AF1 with a single ground state. At about $T_\mathrm{f} \approx 0.06$ the system shows a (quite broad) phase transition to a state with antiferromagnetic order. In contrast to the previous systems, some spins are able to fluctuate much below the ordering temperature resulting in a $1/T$-divergence of the susceptibility $\chi$. We expect that weak interactions will suppress spin fluctuations at low temperatures, so that $\chi$ approaches zero as $T \to 0$. Hence, in some systems $\chi$ shows a peak at low temperatures if spin fluctuations are present.

To study further these fluctuations we compute for each spin $i$ the thermal average of the spin-direction with respect to the ground state $\xi_i$ using  $\langle m_\mathrm{gs}^i \rangle = \langle \mathrm{sign}(M_\mathrm{gs} )\,\sigma_i \xi_i \rangle$. For a paramagnetic state at $T > T_\mathrm{f}$ with no preferred spin direction we expect $\langle m_\mathrm{gs}^i \rangle = 0$. In contrast, we get $\langle m_\mathrm{gs}^i \rangle = 1$ when the system freezes into a single ground state. We find that many of the studied systems have some 'loose spins' that fluctuate strongly even for $T \ll T_\mathrm{f}$. In Fig.~\ref{fig:fluctuations} we show the spin fluctuations $\langle m_\mathrm{gs}^i \rangle $ for the Ammann Beenker tiling in Fig.~\ref{fig:pt-octagonal2} at different temperatures. We find that even at very low temperatures single spins on many of the 8-fold rings and some of the semi-circular clusters are able to fluctuate. Looking at the magnetic interactions this is caused by frustration effects due to presence of ferromagnetic and antiferromagnetic interactions and the rather broad distribution of interaction strengths with only weak inter-cluster interactions. 

Fluctuations due to loose spins are also present in the Ammann Beenker tiling shown in Fig.~\ref{fig:pt-octagonal}, as can be seen from the peak in the susceptibility $\chi$ at low-temperatures. However, this system shows a second type of fluctuations, between the two almost degenerate ground states. This leads to the divergence of the order parameter susceptibility $\chi_\mathrm{op}$ at low temperatures  as can be seen in Fig.~\ref{fig:pt-octagonal}a. This is an artefact of the parallel tempering method which allows fluctuations between different ground states that are not possible with realistic systems. This effect can be observed for all systems in the AF2 class.

\subsection{Finite Size Scaling}
\label{sec:fss}

Finite size scaling is a valuable tool for studying phase transitions in finite systems and extrapolating to the thermodynamic limit. However, it is difficult to employ it for quasicrystals as on the one hand the system size of successive periodic approximants grows fast and on the other hand we require fairly large system sizes to see the effects of the quasiperiodicity. For instance, the size of successive approximants grows with a factor  $\tau = 3+2\sqrt{2}$ for the Ammann Beenker tiling and $\tau = (1+\sqrt{5})/2$ for the Penrose tiling. Further complication are caused by the fluctuations of loose spin, multiple ground states and rather broad transitions in some systems, as all of them lead to a complex temperature behaviour of the relevant observables.

\begin{figure}
    \includegraphics[width =\columnwidth]{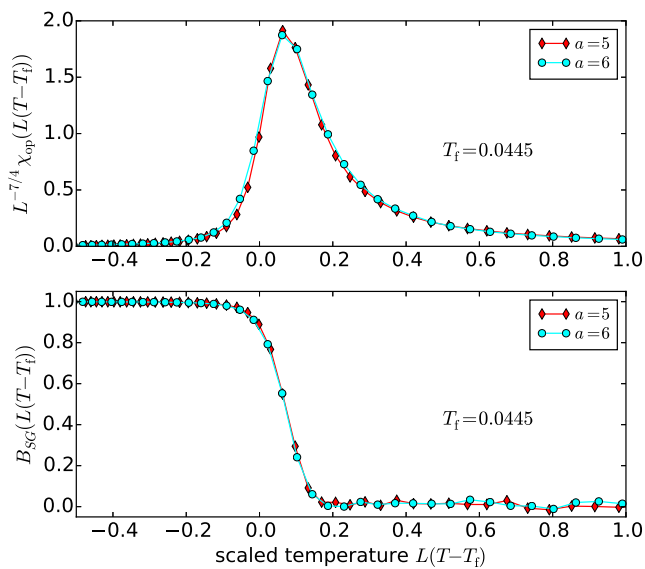}
	\caption{Finite size scaling results of the order parameter susceptibility $\chi_\mathrm{op}$ and the Binder cumulant $B_\mathrm{SG}$ for the Penrose tiling with $E_\mathrm{F} = 0.07$ and $z=5$. Results are shown for the 5th approximant with $N = 375$ and the 6th approximant with $N = 1003$ spins.} 
	\label{fig:fss}
\end{figure}

Luckily, some systems avoid these problems. One example is the Penrose tiling in Fig.~\ref{fig:pt-penrose} with $E_\mathrm{F} = 0.07$ and $z=5$ for which we were able to compute the observables for the 5th and 6th approximants, with 375 and 1003 spins respectively. We also obtained results for smaller approximants but they show quite different transition temperatures because they are too small to adequately describe the quasiperiodic structure. For the two large approximants, we find that the peak of the order parameter susceptibility $\chi_\mathrm{op}$ becomes sharper and the slope of the Binder cumulant $B_\mathrm{SG}$ increases with the system size. 

While in an infinite volume the correlation length $\xi$ diverges near the transition point as $\xi \propto |t|^{-\nu}$ with $t = T - T_\mathrm{f}$, in simulations with finite size $N = L^d$ the system already becomes effectively ordered for correlation lengths $\xi \approx L$. Therefore, the observables can be described by the finite size scaling equations $\chi_\mathrm{op}(L, T) = L^{\gamma/\nu} g_\chi(L^{1/\nu}|t|)$ for the order parameter susceptibility 
and $B_\mathrm{SG} (L, T) = g_B (L^{1/\nu}|t|)$ for the spin glass Binder cumulant.\cite{PhysRevLett.1981.Binder, PhaseTransitions.1996.Cardy} We expect our systems to be in the 2D Ising universality class for which the exact exponents are known to be $\nu = 1$ and $\gamma = 7/4$.\cite{PhysRev.1944.Onsager} To determine an estimate for the critical temperature $T_\mathrm{f}$ we use the Binder cumulant $B_\mathrm{SG}$ which is less sensitive to finite size effects. For the Penrose tiling with $E_\mathrm{F} = 0.07$ and $z=5$, we find an excellent agreement of the scaled Binder cumulant for the two approximants with $T_\mathrm{f} = 0.0445$ (see Fig.~\ref{fig:fss}). Scaling of the order parameter susceptibility is also in very good agreement for this critical temperature.

\subsection{Comparison to Experimental Results}

So far quasiperiodic magnetic order has not been observed in rare earth quasicrystals. \cite{MagProp.2013.Stadnik} Experimental data for many of these materials shows a spin-glass-behaviour with a separation of the field-cooled and zero-field-cooled magnetic susceptibility below a freezing temperature $T_{\rm f}$. \cite{PhilMag.2008.Hippert, MagProp.2013.Stadnik, NatureMat.2013.Goldman} In most cases the Weiss temperature $\Theta$ is negative, indicating predominantly antiferromagnetic interactions. Also  $-\Theta/T_{\rm f}$ is rather large with values of 5 to 10 implying strong frustration. \cite{PhysRevB.1999.Fisher,NatureMat.2013.Goldman}

However, some 1/1 cubic approximants are known to have long-range magnetic order. For instance, the  Cd$_6$R approximants show long-range antiferromagnetic order.\cite{PhysRevB.2010.Tamura} Interestingly, the 1/1 cubic approximant of AuSiGd and AuGeGd are both ferromagnetic below a freezing temperature but at even lower temperatures the AuGeGd approximant alone shows an additional transition to a spin-glass.\cite{JPhysCondMat.2013.Hiroto} According to current structure models, the major difference is some site disorder of the rare earth atoms in AuGeGd which is not present in the AuSiGd approximant.\cite{JPhysCondMat.2013.Gebresenbut} In general, site disorder of the rare earth atoms is very common in quasicrystals. For instance, the structure model for the recently discovered binary quasicrystals CdR also includes site disorder for some of the rare earth positions. \cite{ChemSocRev.2013.Tsai, SciTech.2014.Goldman}  This suggests that the spin freezing observed in rare earth quasicrystals may be attributed to structural disorder.

With respect to observed spin freezing in rare earth quasicrystals, some features are in agreement with canonical spin glasses but there are also important differences. The frequency dependency of the a.c.~magnetic susceptibility and the behaviour of the non-linear susceptibility in i-ZnMgRe are typical for a conventional spin-glass transition. \cite{PhysRevB.1999.Fisher} However, neutron scattering data for i-ZnMgTb and i-ZnMgHo show the formation of short-range order within spin-clusters at about 20K. Although individual spins are frozen within the clusters, the clusters are able to  fluctuate in a rather broad temperature range of about 5.8K $< T <$ 20K. \cite{PhysRevB.2000.Sato, PhysRevB.2006.Sato, PhysRevB.1998.Islam} The latter observation fits well with our results for the formation of strongly coupled spin clusters on certain patterns of the tiling and the fluctuations of these clusters much below the ordering transition being permitted by the weaker inter-cluster interaction.

A natural extension of our model is to include site disorder in the systems and study their magnetic properties. We find in preliminary work that RKKY interactions computed in our model show a high sensitivity to a weak random on-site potential. This also implies that even disorder restricted to non-magnetic sites will influence the couplings between spins.

\section{Conclusion}

We have used Ising spins with RKKY interactions computed from tight-binding models on quasiperiodic tilings as a simple caricature for the magnetic properties of rare-earth quasicrystals. The RKKY interactions were computed with an improved numerical method which directly uses the multifractal electronic eigenstates of the system. For all systems we find the emergence of strongly-coupled spin clusters with weak inter-cluster coupling on certain patterns of the tiling which repeat quasiperiodically. 

Using extensive Monte Carlo simulations, we find that all studied systems show a phase transition at a temperature $T_\mathrm{f}$ at which the global Ising symmetry of the spin model is broken spontaneously. Evidence for this transition includes the temperature-dependence of the order parameter susceptibility, spin glass Binder cumulant, and the overlap distribution. While some systems have large ferromagnetic regions at low temperatures, the majority of systems show a transition to a quasiperiodic Ne\'el state. We demonstrate that the low-temperature state has long-range quasiperiodic magnetic order by analysing the magnetic structure factor. This result is striking taking into account the frustration and quasi-randomness of the magnetic interactions in our quasicrystal model. 

The formation of strongly coupled clusters and their fluctuations even at very low temperatures appears to be consistent with experimental observations of the formation of ordered spin clusters and collective spin fluctuations. In contrast, the nature of the ordering transition in the model shows clear differences to experiments, and it remains an interesting open problem to investigate the origin of the spin freezing found in many of these materials.

This research was supported by a Marie Curie Intra European Fellowship within the 7th European Community Framework Programme and by EPSRC Grant No. EP/I032487/1. 

\bibliographystyle{apsrev4-1}
\bibliography{paper}

\end{document}